\documentclass[preprint,aps,12pt,showpacs]{revtex4}
\usepackage{graphicx,color}
\begin{document}

%=========================================
\newcommand  {\ba} {\begin{eqnarray}}
\newcommand  {\ea} {\end{eqnarray}}
\def\cM{{\cal M}} 
\def\cO{{\cal O}}
\def\cK{{\cal K}}
\def\cS{{\cal S}}
\newcommand{\mh}{m_{h^0}}
\newcommand{\mw}{m_W}
\newcommand{\mz}{m_Z}
\newcommand{\mt}{m_t}
\newcommand{\mb}{m_b}
\newcommand{\be}{\beta}
\newcommand{\al}{\alpha}
\newcommand{\lam}{\lambda}
\newcommand{\no}{\nonumber}
\def\ga{\mathrel{\raise.3ex\hbox{$>$\kern-.75em\lower1ex\hbox{$\sim$}}}}
\def\la{\mathrel{\raise.3ex\hbox{$<$\kern-.75em\lower1ex\hbox{$\sim$}}}}
%==========================================

\title{Probing triple Higgs couplings of the Two Higgs Doublet Model 
at Linear Collider}

\author{Abdesslam Arhrib$^{1,2}$, Rachid Benbrik$^{3}$, Cheng-Wei
  Chiang$^{1,4}$}

\affiliation{$^1$ Department of Physics and Center for Mathematics and
  Theoretical Physics, National Central University, Chungli, Taiwan
  320, R.O.C.}

\affiliation{$^2$ D\'epartement de Math\'ematiques, Facult\'e des
  Sciences et Techniques B.P. 416 Tanger, Morocco.}

\affiliation{$^3$ Department of Physics, Chung Yuan Christian
  University, Chungli, Taiwan 320, R.O.C.}

\affiliation{$^4$ Institute of Physics, Academia Sinica, Taipei,
  Taiwan 115, R.O.C.}

%\date{\today}
\begin{abstract}
  We study double Higgs production at the future Linear Collider in
  the framework of the Two Higgs Doublet Models through the following
  channels: $e^+e^- \to \Phi_i \Phi_j Z$, $\Phi_i=h^0, H^0, A^0,
  H^\pm$.  All these processes are sensitive to triple Higgs
  couplings.  Hence observations of them provide information on the
  triple Higgs couplings that help reconstructing the scalar
  potential.  We also discuss the double Higgs-strahlung $e^+e^- \to
  h^0 h^0 Z$ in the decoupling limit where $h^0$ mimics the SM Higgs
  boson.  The processes $e^+e^- \to h^0 h^0 Z$ and $e^+e^- \to h^0 H^0
  Z$ are also discussed in the fermiophobic limit where distinctive
  signatures such as $4\gamma +X$ , $2\gamma +X$ and $6\gamma +X$ are
  expected in the Type-I Two Higgs Doublet Model.
\end{abstract}


\maketitle
%==============================

\section{Introduction}

In order to establish the Higgs mechanism for the electroweak symmetry
breaking \cite{Higgs:1964pj}, we need to measure the Higgs couplings
to fermions and to gauge boson as well as the self-interaction of
Higgs bosons.  Such measurements, if precise enough, can be helpful in
discriminating between models through their sensitivity to quantum
corrections, particularly in specific scenarios such as the decoupling
limit where Higgs couplings mimic the SM ones.

If electroweak symmetry breaking is achieved by the Higgs mechanism,
it is possible to discover at least one light Higgs boson at the Large
Hadron Collider (LHC).  Such a Higgs boson can be a Standard Model
(SM) one or one of those predicted by various extentions of the SM,
such as the Minimal Supersymmetric Standard Model (MSSM) or Two Higgs
Doublet Model (2HDM).

In case of a discovery of the Higgs boson at the LHC, it may be
possible to measure its couplings to gauge bosons and fermions at a
certain precision \cite{Ruwiedel:2007tv}.  It has been demonstrated in
Ref.~\cite{ILCLHC} that physics at the LHC and the $e^+e^-$
International Linear Collider (ILC) will be complementary to each
other in many respects.  In many cases, the ILC can significantly
improve the LHC measurements.

In recent years there has been growing interest in the study of
extended Higgs sectors with more than one Higgs doublet
\cite{Gun,abdel1,abdel2}.  The simplest extension is the 2HDM; and
such a structure is required for the MSSM.  Models with two (or more)
Higgs doublets predict the existence of charged Higgs bosons.
Therefore, the discovery of them would be the conclusive evidence of
an extended Higgs sector.

For the linear collider, there have been several studies dedicated to
triple Higgs couplings both in the SM and beyond (for a review, see
Ref~\cite{MSSMHIGGS}).  In Refs.~\cite{tripleSM} and
\cite{Osland:1998hv}, the double Higgs-strahlung processes and $WW$
fusion both in the SM and MSSM have been addressed.  In both cases,
the sizes of double Higgs-strahlung processes and $WW$ fusion cross
section are rather small.  It has been shown in Ref.~\cite{miller}
that for the process of $e^+e^- \to ZHH$ with $H\to b\bar{b}$ in the
SM, the irreducible background from electroweak and QCD processes can
be suppressed down to manageable levels by using
kinematics cuts.

At the LHC, the double Higgs production has been studied in
Ref.~\cite{tripleMSSMLHC} with the conclusion that triple Higgs
couplings in the SM and MSSM can be measured provided the background
can be rejected sufficiently well.

In 2HDM's, triple and quartic Higgs couplings have been studied for a
linear collider in Refs.~\cite{Dubinin:1998nt,Dubinin:2002nx,sola}.
In Ref.~\cite{Dubinin:1998nt}, the triple and quartic couplings have
been studied in 2HDM without CP violation.  The analytic expressions
of triple and quartic couplings are given, but numerical evaluations
for the cross section of double Higgs-strahlung are given in the
framework of MSSM.  Ref.~\cite{Dubinin:2002nx} studies the triple
Higgs couplings in the framework of 2HDM with CP violation in the
Higgs sector.  Besides, the numerical analysis for triple Higgs
couplings has been presented in MSSM with explicit CP violation.

Ref.~\cite{sola} concentrates exclusively on the triple Higgs
production of the 2HDM at the linear collider, and the cross sections
are found to be large.  In this paper, we study the double
Higgs-strahlung production $e^+e^- \to \Phi_i \Phi_j Z$, $\Phi_i=h^0,
H^0, A^0, H^\pm$ in the general 2HDM, as measurements of these
processes can shed some light on the 2HDM triple Higgs couplings.  In
addition, we also take into account the perturbativity as well as
vacuum stability constraints on various parameters in the Higgs
potential.  We will show that after imposing those constraints, there
are still large enough cross sections, a few hundred femto-barns (fb)
in some cases, to allow a determination of the 2HDM triple Higgs
couplings.  We will also study some of these processes in the
decoupling limit and in the fermiophobic limit of the so-called type-I
2HDM.

The paper is organized as follows.  In the next section, we give a
short review of 2HDM as well as a rough numerical estimate of the
triple Higgs couplings.  In Section~\ref{sec:Higgsstrahlung}, we
evaluate the double Higgs-strahlung production $e^+e^- \to \Phi_i
\Phi_j Z$ in the 2HDM paying special attention to $e^+e^- \to h^0 h^0
Z$ in the decoupling limit, where $h^0$ mimics the SM Higgs boson, as
well as in the fermiophobic limit.  Our findings are summarized in
Section~\ref{sec:summary}.

\section{Short review of 2HDM and triple Higgs couplings}
\subsection{Short review of 2HDM }

In this section we define the scalar potential to be studied in this
article.  It has been shown in Refs.~\cite{Gun,abdel2} that the most
general 2HDM scalar potential that is both $SU(2)_L\otimes U(1)_Y$ and
CP invariant is given by:
%%%%%%%%%%%%%%%
\ba
V(\Phi_1,\Phi_2)
&=& m^2_1 \Phi^{\dagger}_1\Phi_1+m^2_2 \Phi^{\dagger}_2\Phi_2 -(m^2_{12}
\Phi^{\dagger}_1\Phi_2+{\rm h.c}) +\frac{1}{2} \lam_1 (\Phi^{\dagger}_1\Phi_1)^2 
+\frac{1}{2} \lam_2
(\Phi^{\dagger}_2\Phi_2)^2\nonumber \\ &+&
\lam_3 (\Phi^{\dagger}_1\Phi_1)(\Phi^{\dagger}_2\Phi_2) + \lam_4
(\Phi^{\dagger}_1\Phi_2)(\Phi^{\dagger}_1\Phi_2) +
\frac{1}{2} \lam_5[(\Phi^{\dagger}_1\Phi_2)^2+{\rm h.c.}] ~,
\label{higgspot}
\ea
%%%%%%%%%%%%
where $\Phi_1$ and $\Phi_2$ have weak hypercharge $Y=1$ and vacuum
expectation values (VEV's) $v_1$ and $v_2$, respectively, and
$\lambda_i$ and $m_{12}$ are real-valued parameters.  Note that this
potential violates the discrete symmetry $\Phi_i\to -\Phi_i$ softly by
the dimension-two term $m_{12}^2(\Phi^+_{1}\Phi_{2})$, and has the
same general structure as the scalar potential in the MSSM.

After the electroweak symmetry breaking, the $W^\pm$ and $Z$ gauge
bosons acquire their masses.  Explicitly, three of the eight degrees
of freedom in the two Higgs doublets correspond to the three Goldstone
bosons ($G^\pm$, $G^0$) and the remaining five become physical Higgs
bosons: $h^0$, $H^0$ (CP-even), $A^0$ (CP-odd), and $H^\pm$ with
masses $m_{h^0}$, $m_{H^0}$, $m_{A^0}$, and $m_{H^\pm}$, respectively.

The potential in Eq.~(\ref{higgspot}) has ten parameters (including
$v_1$ and $v_2$).  The parameters $m_1$ and $m_2$ are fixed by the
minimization conditions.  The combination $v^2=v_1^2 + v_2^2$ is fixed
as usual by the electroweak breaking scale through $v^2=(2\sqrt{2}
G_F)^{-1}$.  We are thus left with seven independent parameters;
namely $(\lambda_i)_{i=1,\ldots,5}$, $m_{12}$, and $\tan\beta \equiv
v_2/v_1$.  Equivalently, we can take instead
\ba m_{h^0}\quad ,
\quad m_{H^0} \quad , \quad m_{A^0} \quad , \quad m_{H^\pm} \quad , \quad
\tan\beta \quad , \quad \alpha \quad \rm{and} \quad m_{12}.
\label{parameters} 
\ea
as the seven independent parameters.  The angle $\beta$ diagonalizes
both the CP-odd and charged scalar squared-mass matrices and $\alpha$
diagonalizes the CP-even squared-mass matrix.  One can easily
calculate the physical scalar masses and mixing angles from
Eq.~(\ref{higgspot}) in terms of $\lambda_i$, $m_{12}$ and $v_i$, and
invert them to obtain $\lambda_i$ in terms of physical scalar masses,
$\tan\beta$, $\alpha$, and $m_{12}$ \cite{Gunion:2002zf,am}.

It is straightforward to derive the triple Higgs couplings from the
above scalar potential in Eq.~(\ref{higgspot}).  In the next section,
we list the trilinear scalar self-couplings relevant for our study.
Other relevant couplings involving Higgs boson interactions with gauge
bosons and fermions can be found in Refs.~\cite{Gun,aan}.  We note
that Ref.~\cite{santos} lists the complete Higgs trilinear and quartic
interactions for two types of 6-parameter potentials, referred to as
'Potential A' and 'Potential B'.  Potential A is equivalent to our
potential if $m_{12}\to 0$, and in this limit the Feynman rules in the
next section are in agreement with those in Ref.~\cite{santos}.

There exist three classes of 2HDM's.  The main difference among them
is in the ways they couple the Higgs fields to matter fields.
Assuming natural flavor conservation \cite{Glashow:1976nt}, the two
most popular models are type-I and type-II, denoted by 2HDM-I and
2HDM-II, respectively.  In 2HDM-I, the quarks and leptons couple only
to one of the two Higgs doublet exactly the same as in the SM.  In
2HDM-II, one of the two Higgs fields couples only to down-type quarks
(and charged leptons) and the other one only couples to up-type quarks
(and neutral leptons) in order to avoid the problem of large
flavor-changing neutral currents (FCNC's).  This is also the pattern
found in the MSSM.  The third class of models is type-III, denoted by
2HDM-III.  However, they are generally regarded as problematic as
FCNC's appear at the tree level \cite{rodrigues}.  We note that 2HDM-I
can lead to a fermiophobic Higgs boson $h^0$ in the limit where
$\cos\alpha=0$ \cite{Gun,Gun1}, and the dominant decay mode for the
lightest Higgs boson in this model is $h^0\to \gamma \gamma$ or
$h^0\to WW$, depending on its mass.  In comparison, 2HDM-II does not
have such a feature.

In our analysis we also take into account the following constraints
when the independent parameters are varied.

\begin{itemize}
\item The extra contributions to the $\Delta\rho$ parameter from the
  Higgs scalars \cite{Rhoparam} should not exceed the current limit
  from precision measurements \cite{pdg4}: $ |\Delta\rho| \la
  10^{-3}$.  Such an extra contribution to $\delta\rho$ vanishes in
  the limit $m_{H^\pm}=m_{A^0}$.  To ensure that $\Delta\rho$ be
  within the allowed range, we demand only a small splitting between
  $m_{H^\pm}$ and $m_{A^0}$.

\item From the requirement of perturbativity for the top and bottom
  Yukawa couplings \cite{berger}, $\tan\beta$ is constrained to lie in
  the range $0.3\leq \tan\beta \leq 100$.

\item We note in passing that it has been shown in Ref.~\cite{Oslandk} that the
  latest $B\to X_s \gamma$ branching ratio among others puts a lower bound on
  the charged Higgs mass, $m_{H^\pm} \ga 250$ GeV, in 2HDM-II.  However, the
  conclusion does not apply to 2HDM-I.  Since our analysis is generally valid
  for both 2HDM-I and 2HDM-II, we still consider values of $m_{H^\pm}$ below 250
  GeV for that matter.

\item To constrain the scalar-sector parameters we will use both
  perturbativity constraints on $\lambda_i$ \cite{unit1,abdesunit} as
  well as vacuum stability conditions \cite{vac1,vac2}.  We require
  that all quartic couplings of the scalar potential in
  Eq.~(\ref{higgspot}) remain perturbative: $\lambda_i \leq 8 \pi$ for
  all $i$.  These perturbative constraints are slightly less
  constraining than the full set of unitarity constraints
  \cite{unit1,abdesunit} established using the high energy
  approximation and the equivalence theorem.  For vacuum stability
  conditions, we use \cite{vac1,vac2}:
  \begin{eqnarray}
    \nonumber
    & \lambda_1  > 0\;,\quad\quad \lambda_2 > 0\;, 
    \nonumber\\
    & \sqrt{\lambda_1\lambda_2 } 
    + \lambda_{3}  + {\rm{min}}
    \left( 0 , \lambda_{4}-|\lambda_{5}|
    \right) >0  \label{vac}
  \end{eqnarray} 

\item From the experimental point of view, the combined null searches
  from all four CERN LEP Collaborations give the lower limit
  $m_{H^{\pm}}\ge 78$ GeV at 95\% confidence level (CL), a limit which
  applies to all models satisfying BR($H^{\pm}\to \tau\nu_{\tau}$) +
  BR($H^{\pm}\to cs$)=1 \cite{Abdallah:2003wd}.  Two LEP
  Collaborations (OPAL and DELPHI) have performed a search for a
  charged Higgs decaying to $A W^*$ (assuming $m_{A^0}> 2m_b$) and
  derived limits on the charged Higgs mass \cite{Abdallah:2003wd}
  comparable to those obtained from the search for $H^\pm\to
  cs,\tau\nu$.

  For the neutral Higgs bosons, the OPAL and DELPHI Collaborations
  have put a limit on the masses of $h^0$ and $A^0$ in the 2HDM
  \cite{opal,delphi}.  OPAL concludes that the regions $1 \la m_{h^0}
  \la 55$ GeV and $3 \la m_{A^0} \la 63$ GeV are excluded at 95\% CL,
  independent of $\alpha$ and $\tan\beta$ \cite{opal}.  DELPHI
  Collaboration studies the Higgs to Higgs decay $h^0\to A^0A^0$ in
  $e^+e^-\to h^0Z$ and $h^0A^0$ production and a large portion of the
  $m_{h^0}$-$m_{A^0}$ plane is excluded, depending on the suppression
  factor that enters the cross section formulas \cite{delphi}.  In
  what follows, we will assume that all Higgs masses are greater than
  100 GeV except in the fermiophobic limit where $m_{h^0}$ can be as
  light as 60 GeV.
\end{itemize}
%------------------------------------
\begin{figure}[h!]
\centering
\includegraphics[height=2.2in]{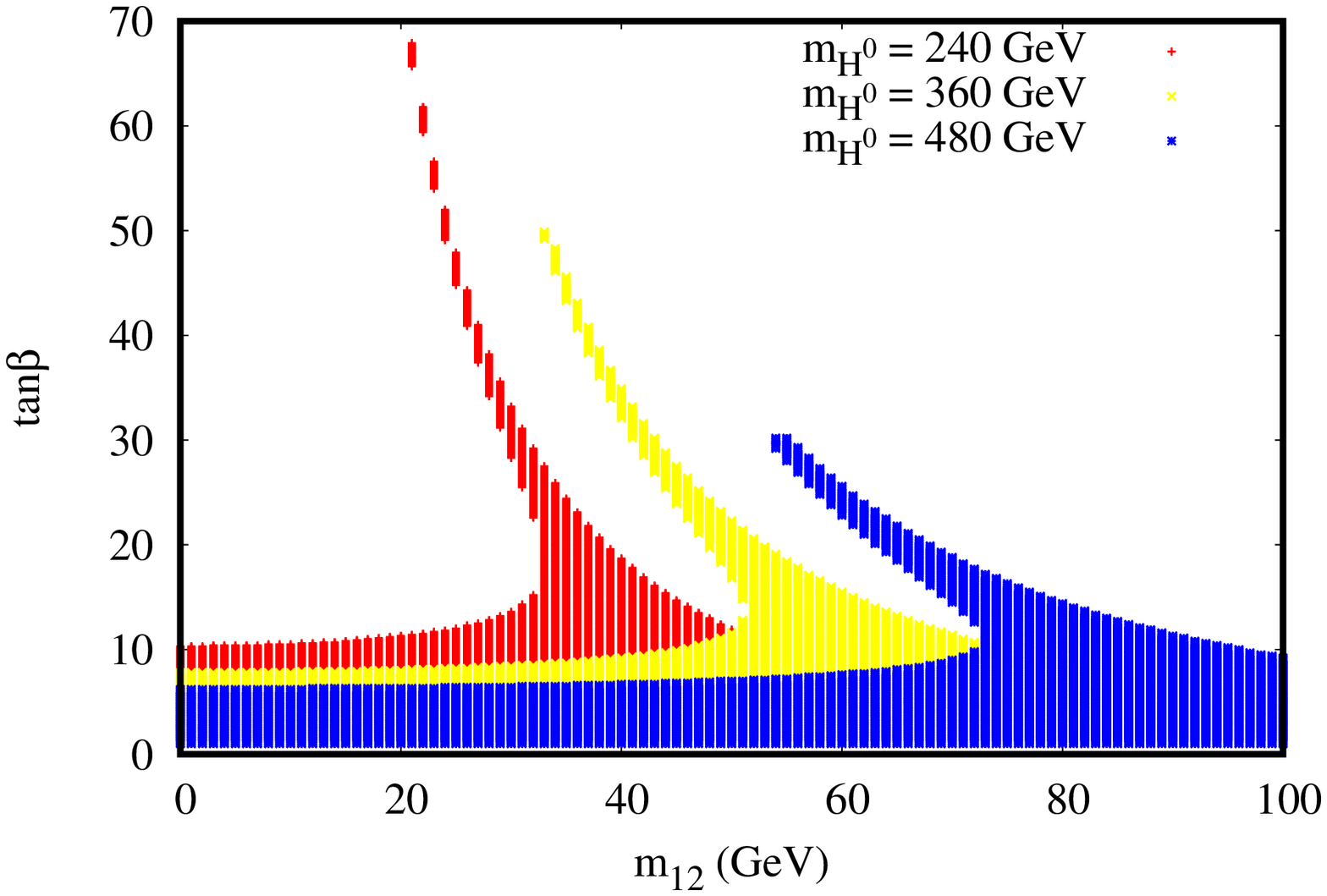}
%\hskip.5cm
\includegraphics[height=2.2in]{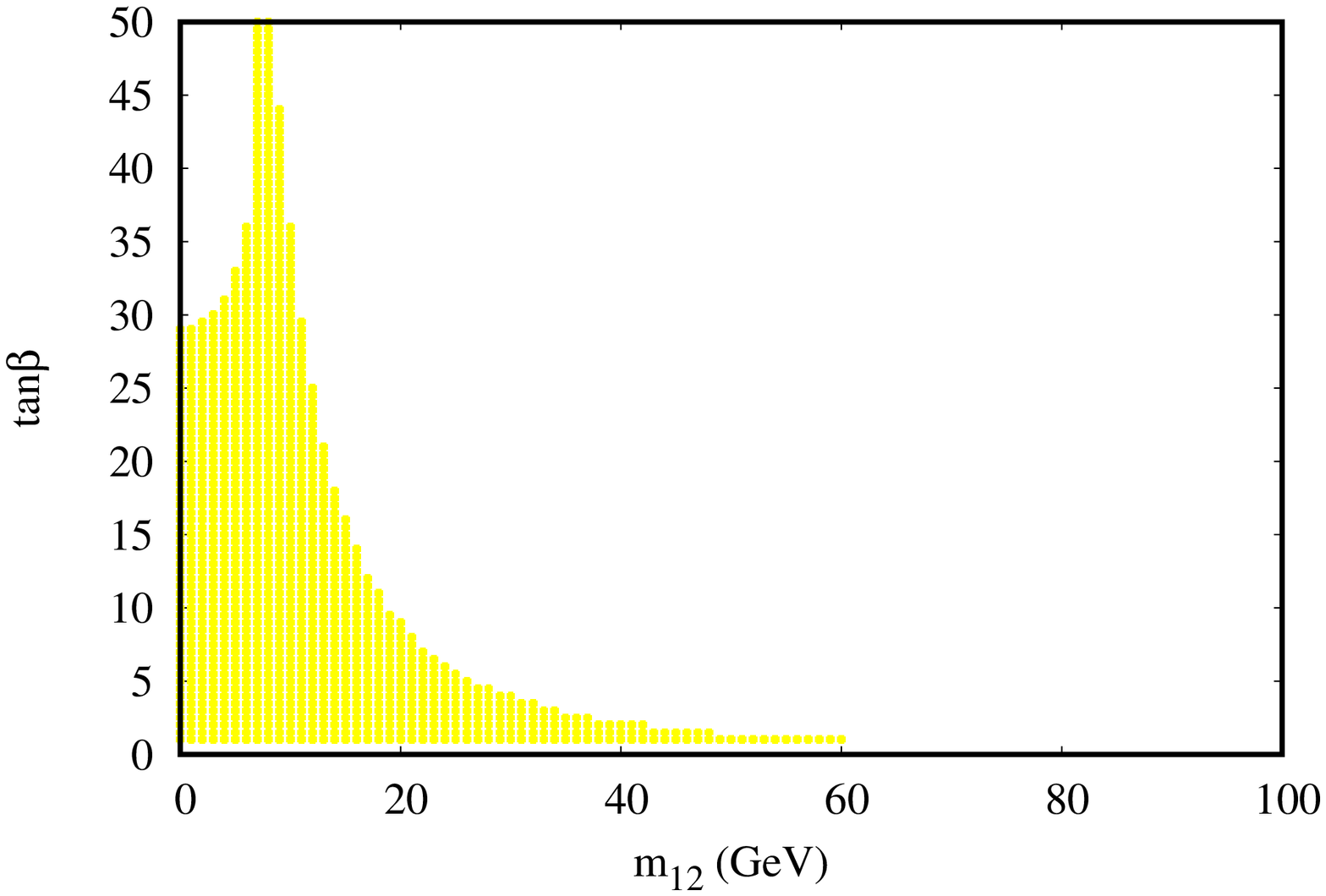}
\caption{Allowed region on the $m_{12}$-$\tan\beta$ plane, taking into account
  perturbativity and vacuum stability constraints.  The common parameters are
  $m_{H^\pm} = 250$ GeV and $m_{A^0} = 150$ GeV.  The left panel uses
  $(\sin\alpha, m_{h^0}) =(0.6, 120 \, {\rm GeV})$ for differents values of
  $m_{H^0}$, and the right panel is for $200 \,{\rm GeV} \le m_{H^0} \le 500$
  GeV and $(\sin\alpha, m_{h^0}) =(1, 60 \, {\rm GeV})$}.
\label{trip0}
\end{figure}
%----------------------------------------

In Fig.~\ref{trip0}, we show the allowed region in the $m_{12}$-$\tan\beta$
plane, taking into account the theoretical constraints mentioned above as well
as $\Delta \rho$.  The surviving parameter region is quite restricted on the
$\tan\beta$-$m_{12}$ plane.  The perturbativity and vacuum stability constraints
together dramatically reduce the allowed parameter space of the model.  In
particular, the perturbativity constraint excludes large values of $\tan\beta$.
In fact, with $\sin\alpha = 0.6$, $m_{h^0} = 120$ GeV and for $m_{H^0} = 240$
GeV to $480$ GeV, the upper bound on $\tan\beta$ decreases from 10 to 5.4 at
$m_{12} = 0$.  Nevertheless, for specific values of $m_{12}\neq 0$ and $m_H$,
one can see that values of $\tan\beta$ as large as 70 can still survive the
perturbativity and vacuum stability constraints.

However, in the fermiophobic limit of 2HDM-I (right panel of Fig.~\ref{trip0})
where $\alpha = \pi/2$, $m_{h^0} = 60$ GeV and $200\, {\rm GeV} \le m_{H^0} \le
500$ GeV, $\tan\beta$ can be as large as about 30 for $m_{12}<$ 15 GeV.  Note
that both plots in Fig.~\ref{trip0} are symmetric under $m_{12} \to -m_{12}$.
For subsequent analyses, we will take $\tan\beta = 10$ as a typical value.

%------------------------------------
\begin{figure}[t!]
\centering
\includegraphics[height=2.2in]{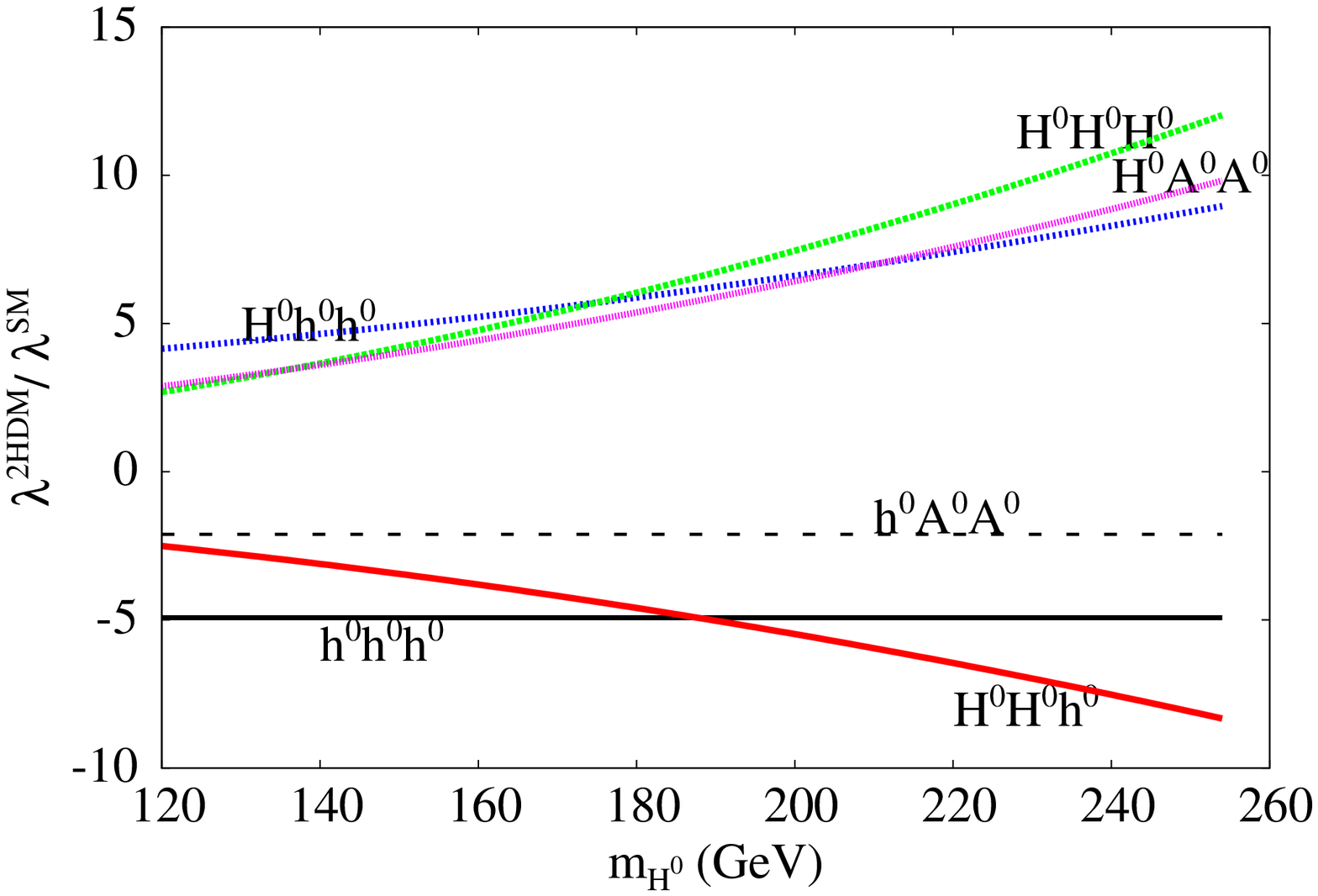}
%\hskip.5cm
\includegraphics[height=2.2in]{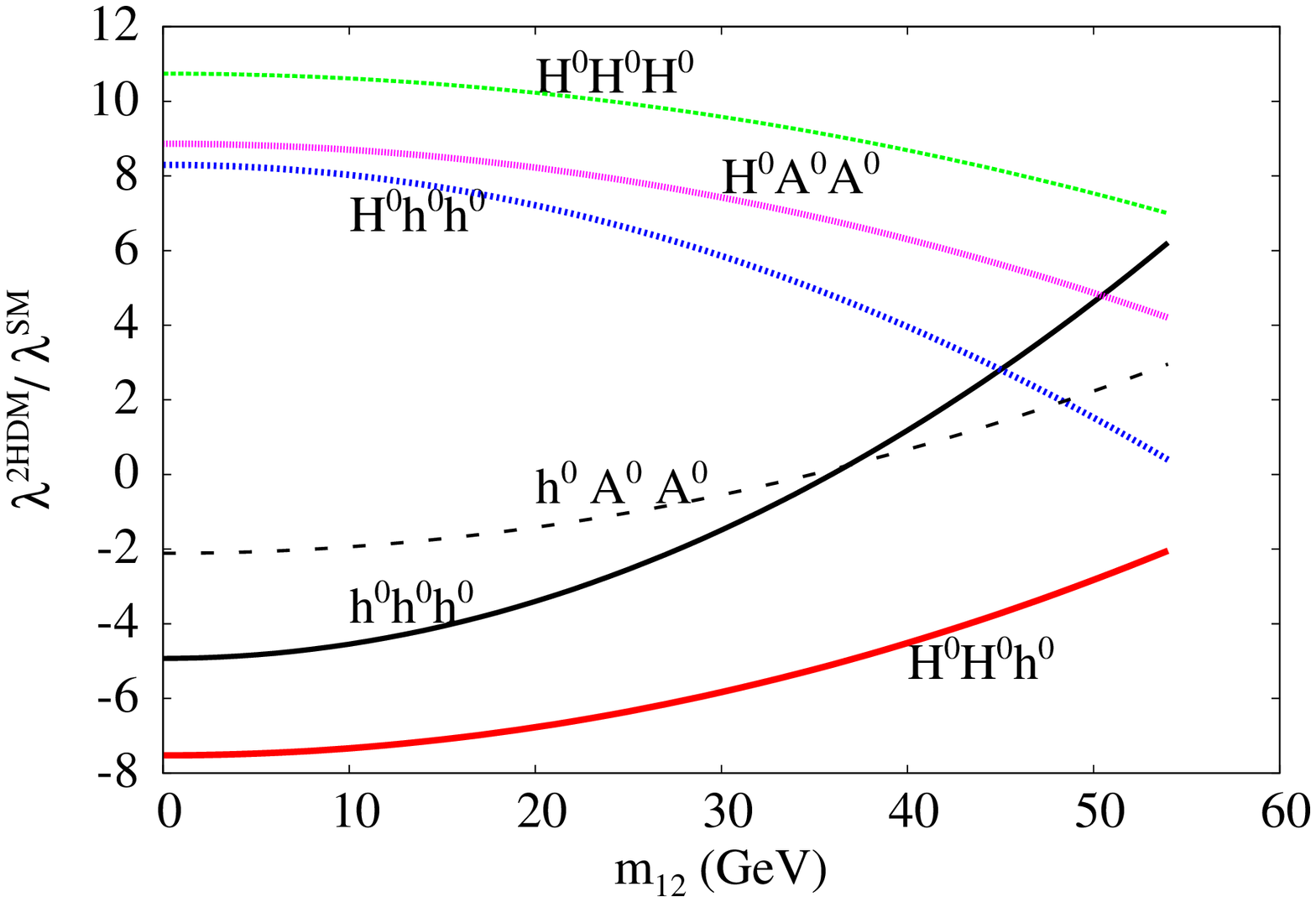}
\caption{The 2HDM tree level self couplings $\lambda_{h_i h_j h_k}^{2HDM}$
  normalized to the SM self coupling $\lambda_{hhh}^{SM}$: (left) as a
  function of $m_{H^0}$ with $m_{12} =0$ GeV, (right) and as a function
  of $m_{12}$ with $m_{H^0}=240 {\rm GeV}$.  The others
  parameters are $m_{h^0} = 120$ GeV, $m_{H^\pm} = 140$ GeV, $m_{A^0}
  = 150$ GeV, $\tan\beta =10$ and $\sin\alpha=0.8$}
\label{trip}
\end{figure}
%----------------------------------------

\subsection{Triple Higgs couplings in 2HDM}

In this section, we will study the behavior of triple Higgs couplings
as a function of the 2HDM parameters $m_{h^0}$, $m_{H^0}$, $m_{A^0}$,
$m_{H^\pm}$, $\tan\beta$, $\alpha$ and $m_{12}$.  These couplings are
given by:
\ba
\lambda_{h^0h^0h^0}^{SM} & = & \frac{-3em_{h^0}^2}{2 m_W s_W}\label{hhhsm}\\
\lambda_{h^0h^0h^0}^{2HDM} &=& \frac{-3e}{m_W s_W s^2_{2\be}}\bigg[(c_\be c^3_\al - s_\be
  s^3_\al)s_{2\be} m^2_{h^0} - c^2_{\be-\al} c_{\be + \al} m^2_{12}\bigg]
\label{lll}\\
\lambda_{H^0H^0H^0}^{2HDM} &=& \frac{-3e}{m_W s_W s^2_{2\be}}\bigg[(c_\be c^3_\al - s_\be
  s^3_\al)s_{2\be} m^2_{H^0} - s^2_{\be-\al} s_{\be + \al} m^2_{12}\bigg] 
\label{HHH}  \\
\lambda_{H^0h^0h^0}^{2HDM} &=& -\frac{1}{2}\frac{e c_{\be-\al}}{m_W s_W s^2_{2\be}}\bigg[
  (2 m^2_{h^0} + m^2_{H^0}) s_{2\al} s_{2\be} - (3 s_{2\al}-s_{2\be})
  m^2_{12}\bigg]\label{hll} \\
\lambda_{H^0H^0h^0}^{2HDM} &=& \frac{1}{2}\frac{e s_{\be-\al}}{m_W s_W s^2_{2\be}}\bigg[
  (m^2_{h^0} + 2 m^2_{H^0}) s_{2\al} s_{2\be} - (3 s_{2\al}+s_{2\be})
  m^2_{12}\bigg] \label{hhl}\\
\lambda_{A^0A^0h^0}^{2HDM} &=& \frac{-e}{m_W s_W s^2_{2\be}}\bigg[(c_\al c^3_\be - s_\al
  s^3_\be)s_{2\be} m^2_{h^0} - c_{\be+\al} m^2_{12}  + s^2_{2\be}
  s_{\be-\al} m^2_{A^0} \bigg]\label{aal}\\
\lambda_{A^0A^0H^0}^{2HDM} &=& \frac{-e}{m_W s_W s^2_{2\be}}\bigg[(s_\al c^3_\be + c_\al
  s^3_\be)s_{2\be} m^2_{H^0} - s_{\be+\al}m^2_{12} +s^2_{2\be}
  c_{\be-\al} m^2_{A^0} \bigg]\label{aah}\\
\lambda_{A^0G^0h^0}^{2HDM} &=& \frac{1}{2}\frac{e c_{\be-\al}}{m_W s_W}\bigg( m^2_{A^0}
- m^2_{h^0}\bigg)\label{agl}\\
\lambda_{A^0G^0H^0}^{2HDM} &=& -\frac{1}{2}\frac{e s_{\be-\al}}{m_W s_W}\bigg( m^2_{A^0}
- m^2_{H^0}\bigg)\label{agh}\\
\lambda_{H^\pm H^\mp h^0}^{2HDM}&=& \frac{e}{m_W s_W s^2_{2\be}}\bigg[(s_\al s^3_\be - c_\al
  c^3_\be)s_{2\be} m^2_{h^0} + c_{\be+\al} m^2_{12} - s^2_{2\be}
  s_{\be-\al} m^2_{H^\pm} \bigg]\label{lhphm}\\
\lambda_{H^\pm H^\mp H^0}^{2HDM} &=& \frac{-e}{m_W s_W s^2_{2\be}}\bigg[(s_\al c^3_\be + s_\al
  s^3_\be)s_{2\be} m^2_{H^0} - s_{\be+\al} m^2_{12} + s^2_{2\be}
  c_{\be-\al} m^2_{H^\pm} \bigg]\label{hhphm}
\ea
where $\lambda_{h^0h^0h^0}^{SM}$ is the SM triple Higgs coupling.
Here we use the short-hand notations: $s_W = \sin\theta_W$ with
$\theta_W$ being the weak mixing angle, and $s_\phi$ and $c_\phi$
denote respectively $\sin\phi$ and $\cos\phi$ for $\phi$ being various
linear combinations of $\alpha$ and $\beta$.

As we can see from Eqs.~(\ref{lll})-(\ref{hhphm}), all triple Higgs
couplings have some quadratic dependence on the physical masses
$m_{\Phi}$ and soft breaking term $m_{12}$.  These couplings also
depend strongly on $\tan\beta$ and $\alpha$.  In the present study, we
will show that varying the scalar parameters within the allowed range
can still make the triple Higgs couplings larger than the SM triple
Higgs coupling $\lambda_{h^0h^0h^0}^{SM}=-3em_{h^0}^2/(2 m_W s_W)$ by
several orders of magnitude.  These 2HDM triple Higgs couplings can
also be large compared to the MSSM triple Higgs couplings.  The reason
for this is that in the MSSM, supersymmetry imposes restrictions on
the quartic couplings $\lambda_i$ by relating them to the gauge
couplings \cite{abdel2,tripleSM}.

Taking into account all the previous theoretical and experimental
constraints listed in the above subsection, we give here the sizes of
the triple Higgs couplings involved in the double Higgs-strahlung
production.
 
%------------------------------------
\begin{figure}[t!]
\centering
\includegraphics[height=2.2in]{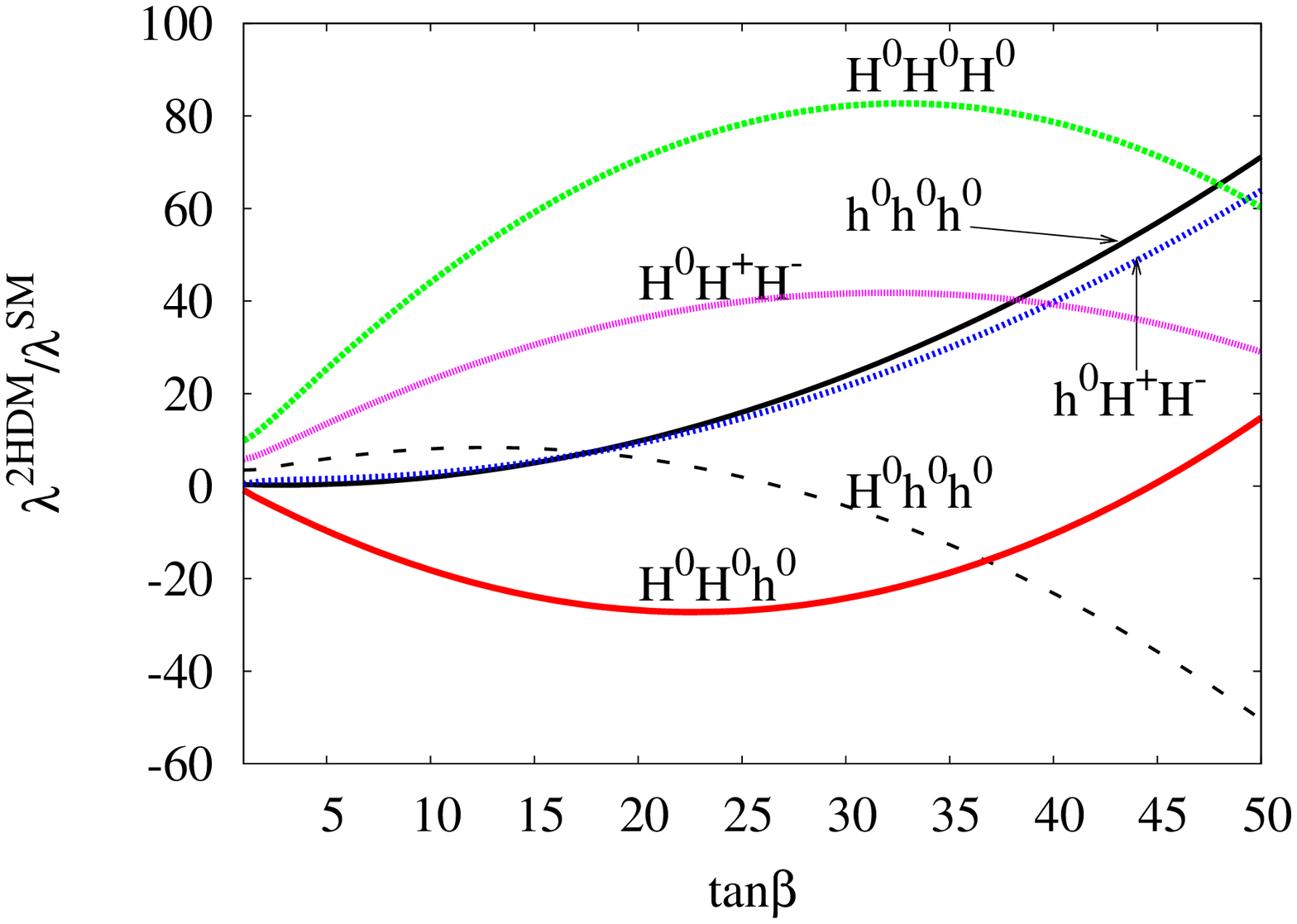}
\includegraphics[height=2.2in]{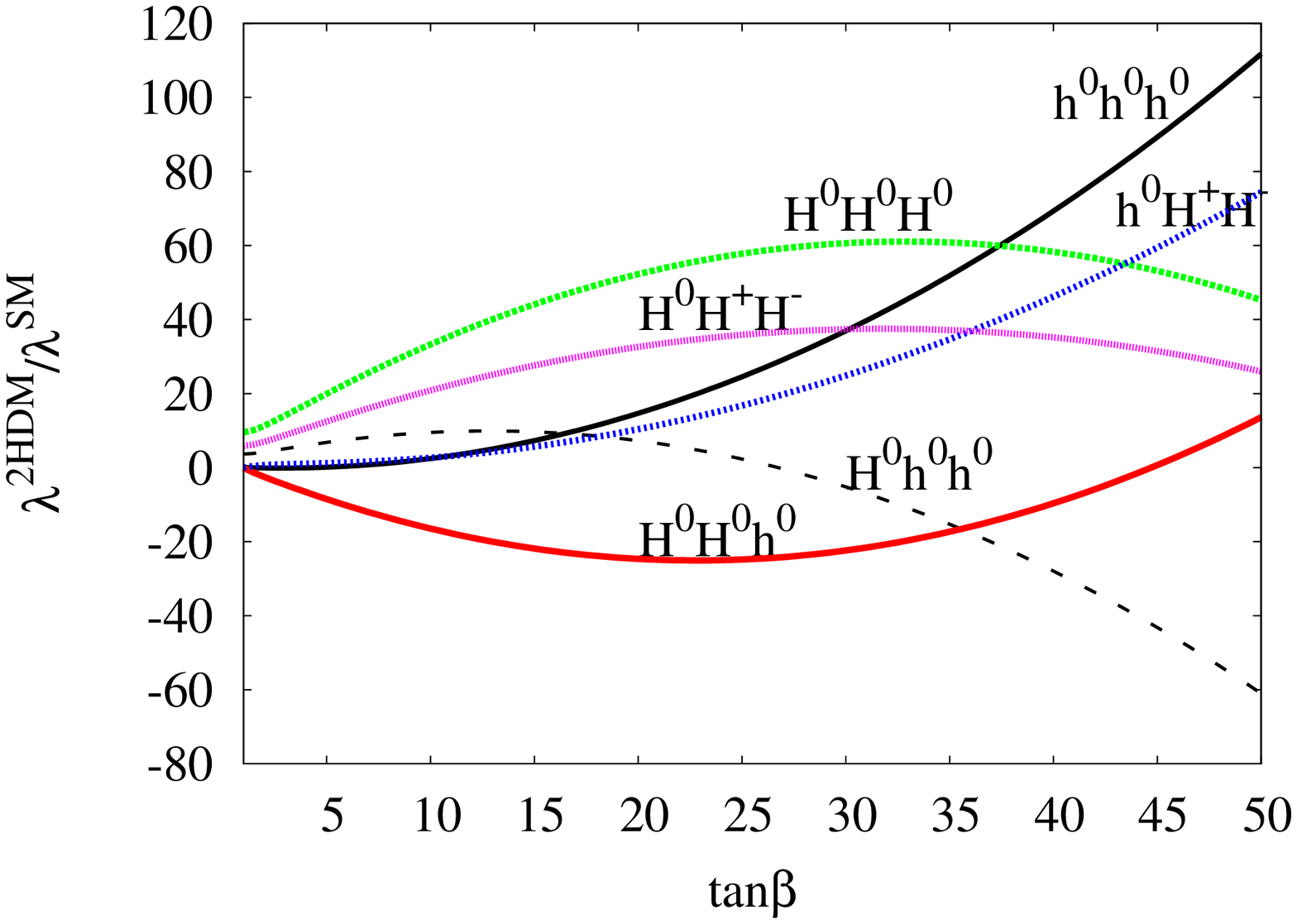}
\caption{The 2HDM tree level self couplings $\lambda_{h_i h_j
    h_k}^{2HDM}$ normalized to the SM self coupling
  $\lambda_{hhh}^{SM}$ as a function of $\tan\beta$.  The plots use
  $m_{h^0} = 120$ GeV, $m_{H^0} = 370$ GeV, $m_{A^0}
  = 150$ GeV, $m_{H^\pm} = 250$ GeV, $m_{12}=46$ GeV, $\sin\alpha
  = 0.6$ (left panel) and $\sin\alpha=0.7$ (right panel).}
\label{trip2}
\end{figure}
%----------------------------------------

In Fig.~\ref{trip}, we illustrate the sizes of the triple Higgs
couplings as a function of CP-even Higgs mass $m_{H^0}$ (left plot)
and as a function of the discrete symmetry breaking parameter $m_{12}$
(right plot).  One can see from these plots that the sizes of triple
Higgs couplings normalized to the SM triple Higgs coupling lie between
$-10$ and 12 for all allowed values of $m_{H^0}$ and $m_{12}$.

If the splitting between $H^\pm$ and $A^0$ is small, it follows from
Eqs.~(\ref{aal}) and (\ref{lhphm}) [Eqs.~(\ref{aah}) and
(\ref{hhphm})] the couplings $\lambda_{h^0A^0A^0}$ and
$\lambda_{h^0H^+H^-}$ [$\lambda_{H^0A^0A^0}$ and
$\lambda_{H^0H^+H^-}$] are almost equal.  The couplings
$\lambda_{h^0H^+H^-}$ and $\lambda_{H^0H^+H^-}$ are thus not shown.
It is clear from Eqs.~(\ref{lll}) and (\ref{aah}) that
$\lambda_{h^0h^0h^0}$ and $\lambda_{h^0A^0A^0}$ have no $m_{H^0}$
dependence.  That is why $\lambda_{h^0h^0h^0}$ and
$\lambda_{h^0A^0A^0}$ exhibit no variation over $m_{H^0}$.  For the
other couplings, the variation over $m_{H^0}$ can change the size of
these couplings by about one order of magnitude.

The dependence of the triple Higgs couplings on $m_{12}$ is also
important (see the right plot of Fig.~\ref{trip}).  However, due to
the perturbativity and vacuum stability conditions, $m_{12}$ is
constrained to be less than 55 GeV.  It is interesting to note that
the couplings $\lambda_{h^0h^0h^0}$ and $\lambda_{H^0h^0h^0}$
contributing to $e^+e^-\to Zh^0h^0$ enter with opposite signs and may
have a destructive interference in the amplitude.  The same
observation holds for $\lambda_{h^0A^0A^0}$, $\lambda_{H^0A^0A^0}$,
$\lambda_{h^0H^+H^-}$ and $\lambda_{H^0H^+H^-}$.

In Fig.~\ref{trip2}, we show the variation of triple Higgs couplings
as a function of $\tan\beta$. We fix $m_{12}=46$ GeV, $m_{h^0} = 120$
GeV, $m_{H^0} = 370$ GeV, $m_{A^0} = 150$ GeV, $m_{H^\pm} = 250$ GeV,
and $\sin\alpha = 0.6$ for left panel and $\sin\alpha=0.7$ for right
panel.  Almost all the triple Higgs couplings are enhanced in the
large $\tan\beta$ limit.  The triple Higgs couplings
$\lambda_{H^0H^0H^0}$, $\lambda_{h^0h^0h^0}$, $\lambda_{H^0h^0h^0}$
and $\lambda_{h^0H^+H^-}$ can be more than 50 times larger than the SM
triple coupling.  This $\tan\beta$ effect has been first noticed in
Refs.~\cite{sola,sola1}.  It is also worth pointing out that some of
the 2HDM couplings $\lambda_{h_ih_jh_k}$ can flip sign as one varies
$m_{12}$ and/or $\tan\beta$.

Note that only in Fig.~\ref{trip2} we ignore perturbativity and vacuum
stability constraints.  In fact, in the left pannel (right panel),
perturbativity constraints are violated for $6\la \tan\beta \la 49$
($7.5\la \tan\beta \la 48$) while vacuum stability constraints are
violated for $43\la \tan\beta \la 49$ ($36\la \tan\beta \la 50$).

%===================================================
\begin{figure}[t!]
\centering
\includegraphics[height=4in]{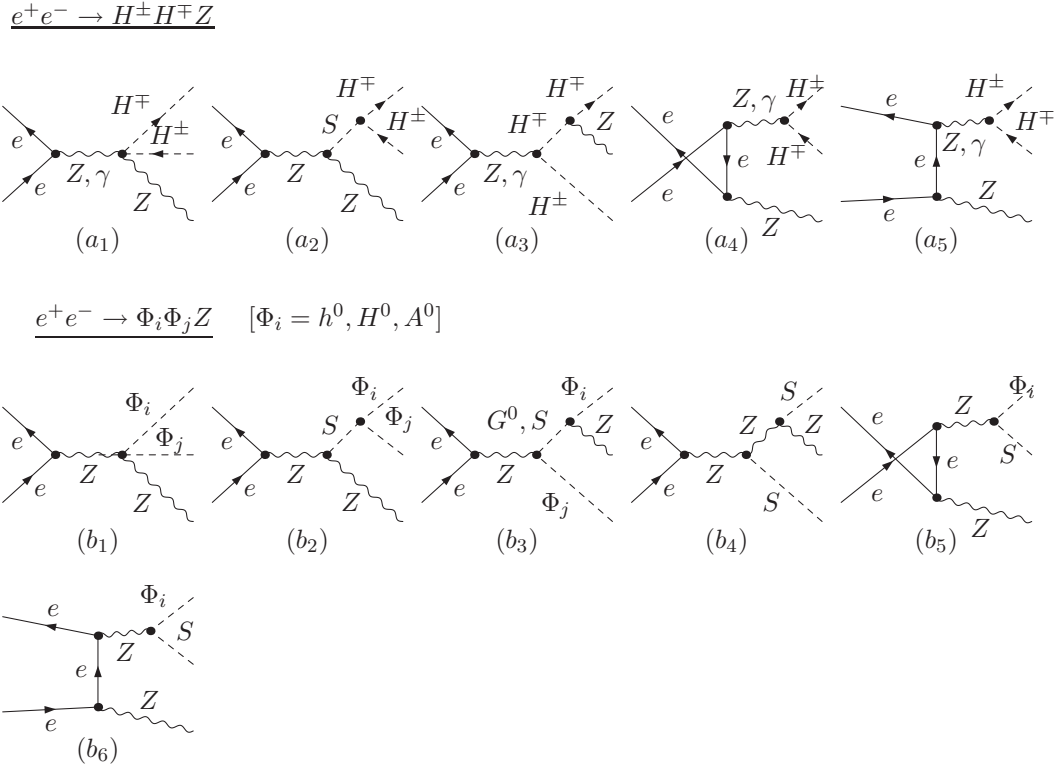}
\caption{Double charged and neutral Higgs-strahlung production in the
  2HDM at $e^+ e^-$ linear colliders with $S=h^0$ or $H^0$ .}
\label{fig:eehhz-diagrams}
\end{figure}
%----------------------------
\section{Probing triple couplings from double Higgs-strahlung
  processes
\label{sec:Higgsstrahlung}}

\subsection{$e^+e^- \to \Phi_i  \Phi_j Z$}

In this section we will cover the following processes: $e^+e^- \to H^+
H^- Z$, $e^+e^- \to h^0 h^0 Z$, $e^+e^- \to h^0 H^0 Z$, $e^+e^- \to
H^0 H^0 Z$ , and $e^+e^- \to A^0 A^0 Z$.  The two processes $e^+e^-
\to h^0 A^0 Z$ and $e^+e^- \to H^0 A^0 Z$ are not sensitive to any
triple Higgs couplings; they proceed only through gauge couplings and
will not be calculated here.

Measurements of those processes will give some information on the involved
 triple Higgs couplings.  The complete calculation is done with the
packages FeynArts \cite{feynarts}, FormCalc \cite{formcalc}, and
LoopTools \cite{looptools}.  In these $2\to 3$ processes, a width
for the internal Higgs exchange is needed to stabilize the phase space
integration.  Such a width is introduced and computed at the tree
level.  The numerical evaluations of the integration over $2\to 3$
phase space is done with the help of CUBA library \cite{cuba}.

%-------------------------------------------------------------------
\begin{figure}[t!]
\centering
\includegraphics[height=2.2in]{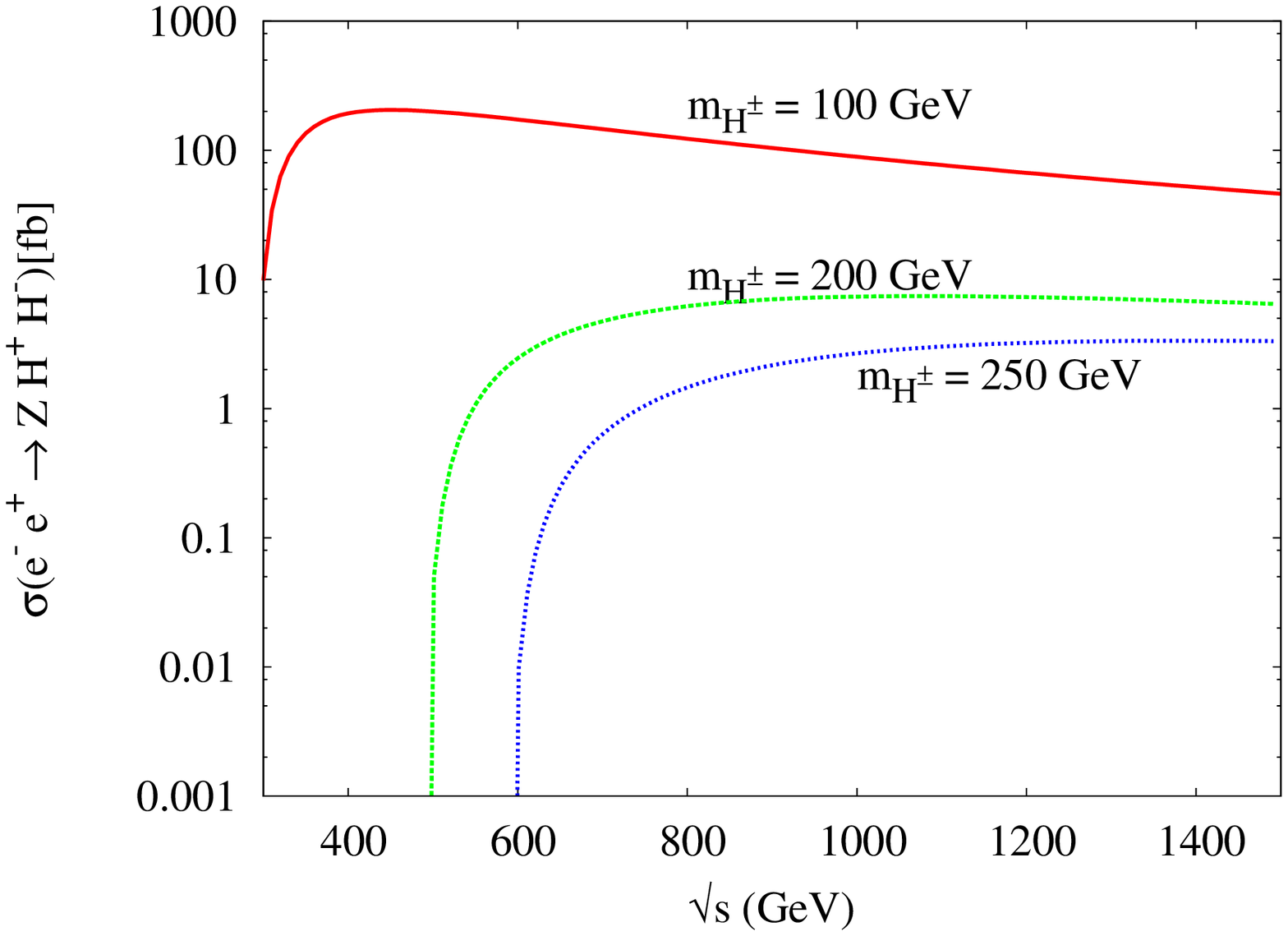}
%\hskip.5cm
\includegraphics[height=2.2in]{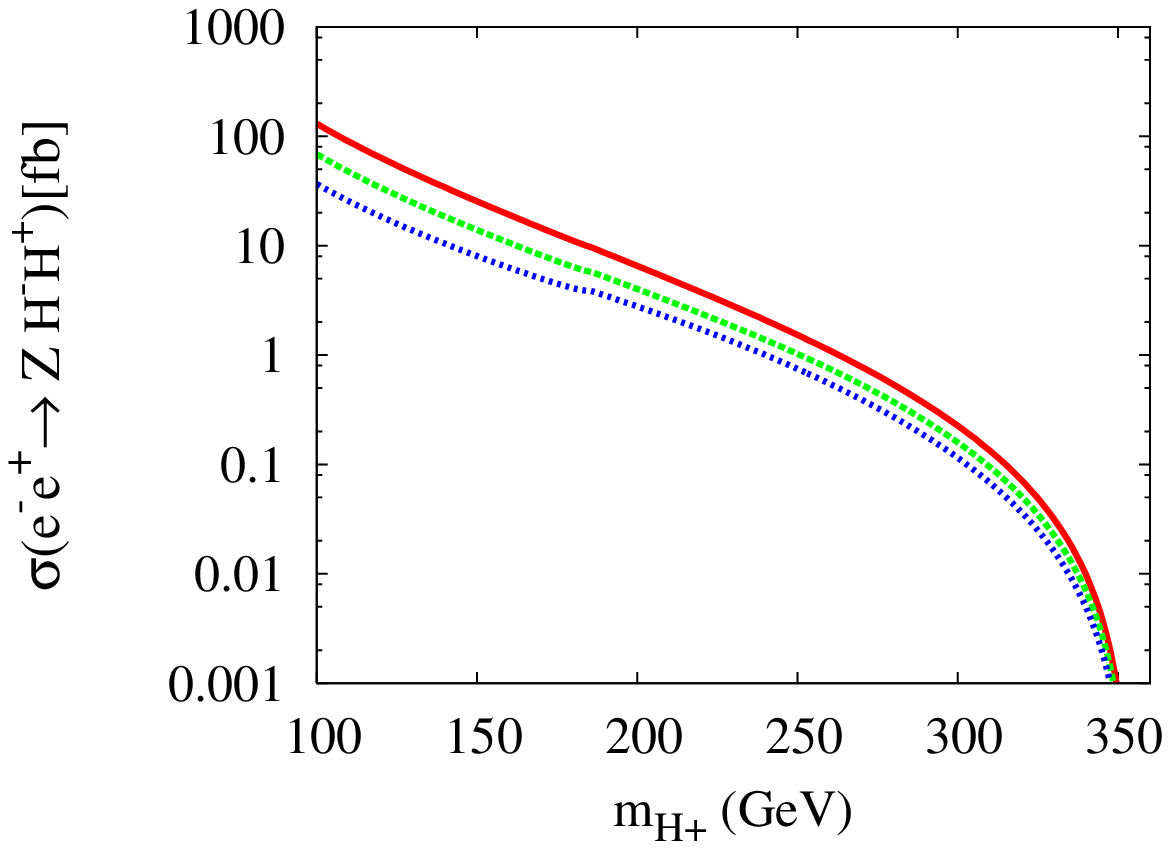}
\caption {$\sigma(e^+ e^- \to H^+H^- Z)$ in units of fb as a function
  of center-of-mass energy (left) for several values of $m_{H^\pm}$
  and as a function $m_{H^\pm}$ (right) for $\sqrt{s}=800$ GeV.
  Regarding the other parameters, the left plot uses $m_{h^0} = 120$
  GeV, $m_{H^0} = 370$ GeV, $m_{12}=45$ GeV, $m_{A^0}=150$ GeV,
  $\tan\beta=53$ and $\cos(\beta-\alpha)=0.5$.  From top to bottom in
  the right plot, ($\tan\beta$, $m_{12}$)=(55,44 GeV), (40,52 GeV),
  and (30, 60 GeV).}
\label{plot1}
\end{figure}
%----------------------------------------

Before discussing our numerical results, it is worth pointing out that
the following results are valid for both 2HDM-I and 2HDM-II since they
involve only the Higgs coupling and the gauge coupling to Higgs or to
$e^+e^-$.

Feynman diagrams for the first 
process $e^+e^- \to H^+ H^- Z$ are depicted in the first set
of Fig.~\ref{fig:eehhz-diagrams}.  Only diagram $(a_2)$ is sensitive
to the triple Higgs couplings $\lambda_{h^0H^+H^-}$ and
$\lambda_{H^0H^+H^-}$.  The other diagrams proceed through gauge
interactions only.  Similarly, Feynman diagrams for the processes 
$e^+e^- \to \Phi_i\Phi_j Z$
with $\Phi_i=h^0, H^0, A^0$ are depicted in the second set of
Fig.~\ref{fig:eehhz-diagrams}.  As we can see, only diagram $(b_2)$
has a triple Higgs couplings dependence.  The other diagrams depend
only on gauge couplings.

To illustrate the size of $e^+ e^- \to H^+H^- Z$ cross section, we
show in Fig.~\ref{plot1} $\sigma(e^+ e^- \to H^+H^- Z)$ as a function
of center-of-mass energy for $m_{H^\pm}=100, 150, 200$ GeV.  The
cross section exceeds $100$ fb for light charged Higgs mass of 100
GeV.  For moderate charged Higgs mass of 200 GeV, the cross section is
still of the order of a few fb.

In the right panel of Fig.~\ref{plot1}, we illustrate the sensitivity
of the cross section to $m_{H^\pm}$ for a fixed center-of-mass energy
of $800$ GeV.  By varying $m_{H^\pm}$ from $100$ to $350$ GeV, the
cross section is suppressed by several orders of magnitude.  However,
for a charged Higgs mass in the range $100$ to $300$ GeV, it is still
possible to have a total cross section larger than $0.1$ fb.  This can
lead to 100 raw events for the planned luminosity of $1000$ fb$^{-1}$.

The collider signature for the $e^+ e^- \to H^+H^- Z$ process depends
on how $H^\pm$ decays.  Below the top-bottom threshold and before the
charged Higgs to neutral Higgs decays $H^\pm \to S W^\pm$
($S=h^0,A^0$) are open, the signature would be $Z\tau^+\tau^- + {\not
  \hspace{-3pt} E}_T$ or $Z \, c \, \bar{s} \, c \, \bar{s}$.  Once
the charged Higgs to neutral Higgs decays $H^\pm \to W^\pm S$ become
open, one would get the $ZW^+W^-SS$ final states with $S$ decaying
either to $b\bar{b}$ or $\tau^+\tau^-$.  Above the top-bottom
threshold, the signature would be $Z \, t \, \bar{t} \, b \, \bar{b}$
if $Br(H^\pm \to t \bar{b})$ dominates or $ZW^+W^-SS$ if $Br(H^\pm \to
W^\pm S)$ dominates.

Let us now turn to double Higgs-strahlung processes $e^+e^- \to
\Phi_i\Phi_j Z$, with $\Phi_{i,j}$ being neutral Higgs bosons $h^0$,
$H^0$, and $A^0$.  As noted in the previous section, the
$\Phi_iH^+H^-$ and $\Phi_iA^0A^0$ couplings ($\Phi_i=h^0, H^0$) are
almost identical for not too large mass splitting between charged and
CP-odd Higgs.  As we will see, the total cross section of $e^+e^- \to
Z A^0A^0$ has similar behavior and size to the total cross section of
$e^+e^- \to H^+ H^- Z$ for $m_{A^0}$ and $m_{H^\pm}$ of about the same
magnitude.

%%%%%%%%%%%%%%%%%%%%%%%%%%%%%%%%%%%%%%%
\begin{figure}[t!]
\centering
\includegraphics[height=2.2in]{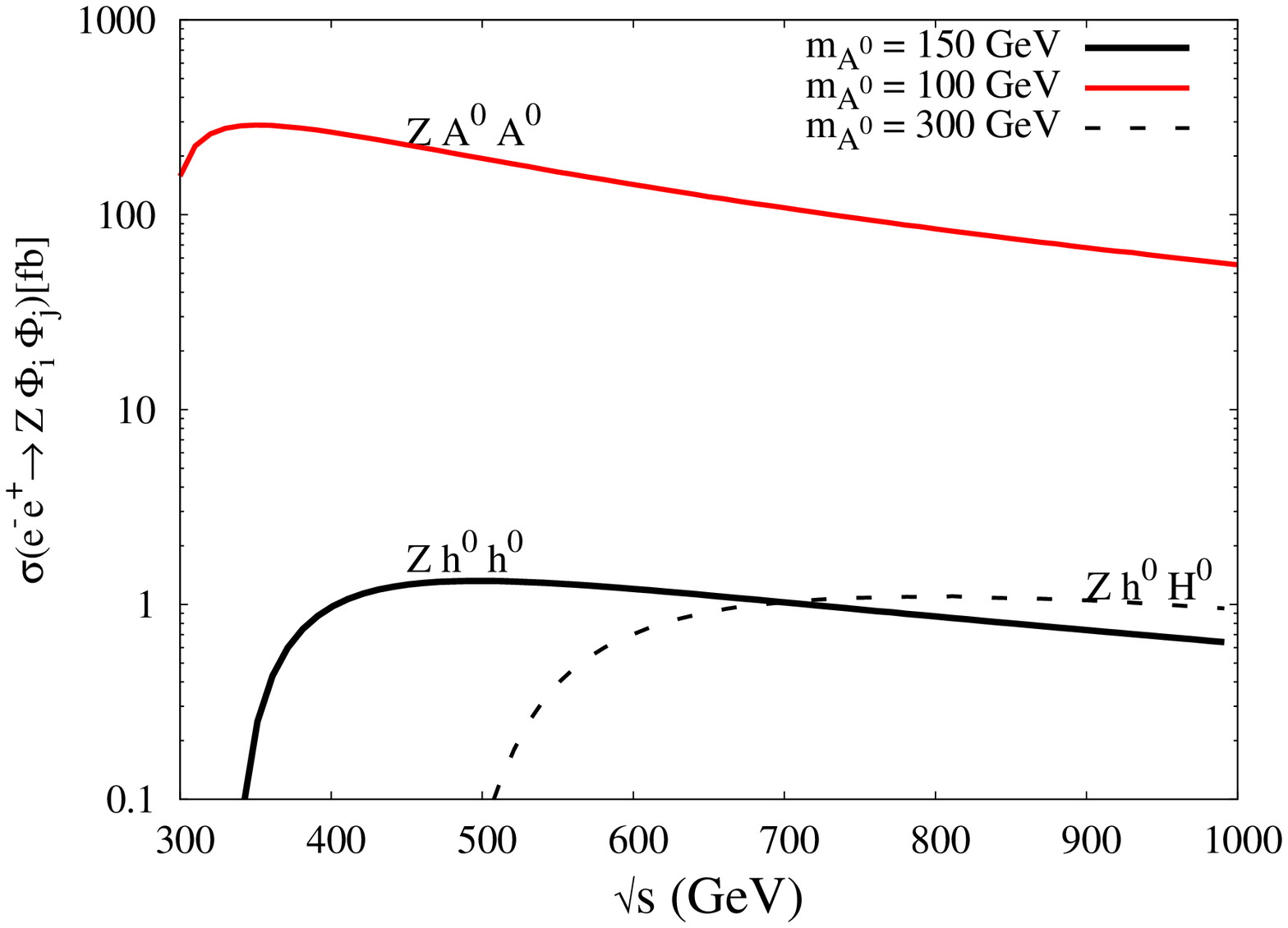}
%\hskip.5cm
\includegraphics[height=2.2in]{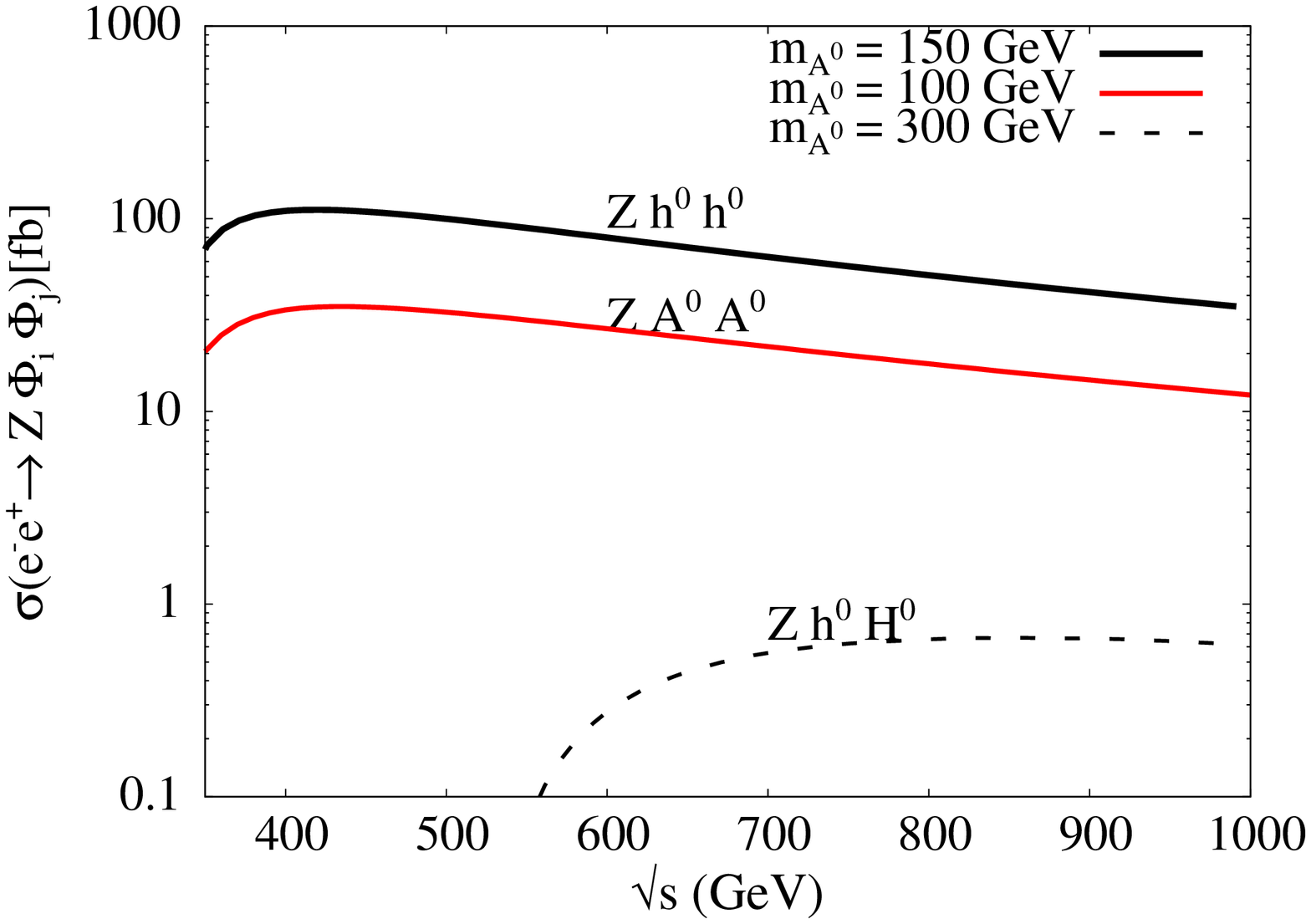}
\caption{Total tree level cross section of $e^+ e^- \to Z \Phi_i
  \Phi_i$ in 2HDM as a function of $\sqrt{s}$ with $m_{h^0} = 120$
  GeV, $m_{H^\pm} = 140$ GeV, $m_{12} = 0$ GeV, $\tan\beta = 10$ for
  different values of $m_{A^0}$.  The left plot uses $m_{H^0}=200$ GeV
  and $\sin\alpha=0.6$.  The right plot uses $m_{H^0}=240$ GeV and
  $\sin\alpha=0.8$.}
\label{ffig1}
\end{figure}
%--------------------------------------

In Fig.~\ref{ffig1}, we show the total cross sections for the neutral
modes $Zh^0h^0$, $Zh^0H^0$, $ZH^0H^0$ and $ZA^0A^0$ as a function of
the center-of-mass energy $\sqrt{s}$ for different values of
$m_{A^0}$.  As explained in the figure caption, we select specific
$\sin\alpha$ values to optimize the total cross sections.  The
behavior of the cross section for those neutral modes is similar to
that of $e^+ e^- \to ZH^+H^-$.  The reasons are that for both
processes we have similar topologies in the contributing Feynman
diagrams and that the triple Higgs couplings have a similar magnitude.
For the $Zh^0h^0$ and $ZA^0A^0$ modes, the total cross section is
maximized for $\sqrt{s}\approx 350-400$ GeV.  In the case of $Zh^0h^0$
with $m_{h^0}=120$ GeV, the maximum of the total cross section is
reached for $m_{A^0}=150$ GeV, $m_{H^0}=240$ GeV, and $\sin\alpha=0.8$
(right panel), and is slightly larger than 100 fb.  This enhancement
for $e^+ e^- \to Zh^0h^0$ is due to the threshold effect ($H^0\to
h^0h^0$ is open).  In the $ZA^0A^0$ mode, the cross section is maximal
for $m_{A^0}=150$ GeV, $m_{H^0}=200$ GeV, and $\sin\alpha=0.6$ and is
about 300 fb.  The cross section of the heavy mode $ZH^0H^0$ is very
small and is not shown; whereas the cross section of the mixed mode
$ZH^0h^0$ is mild, $\approx 1$ fb, in both cases.

\begin{figure}[t!]
\centering
\includegraphics[height=2.2in]{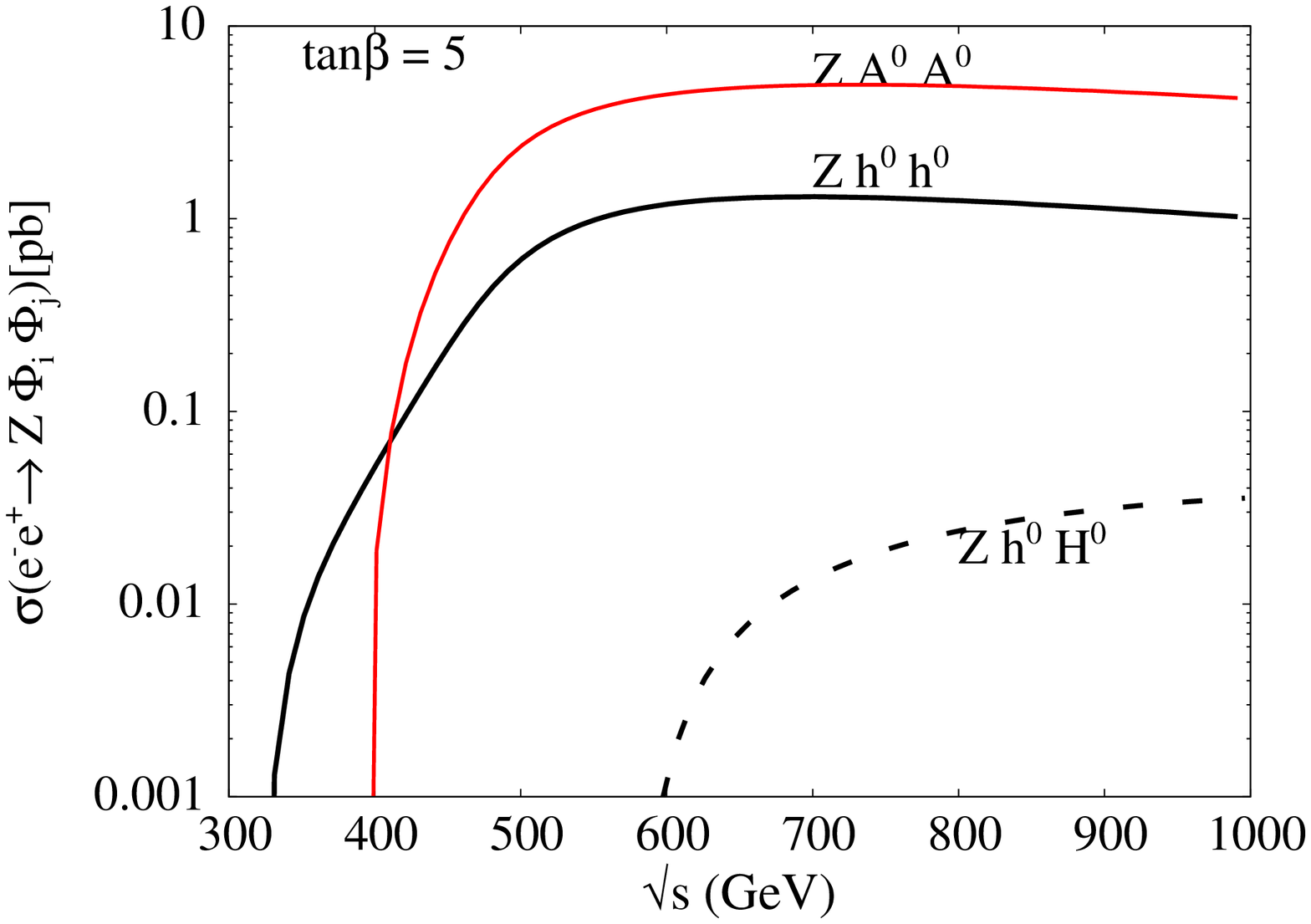}
%\hskip.5cm
\includegraphics[height=2.2in]{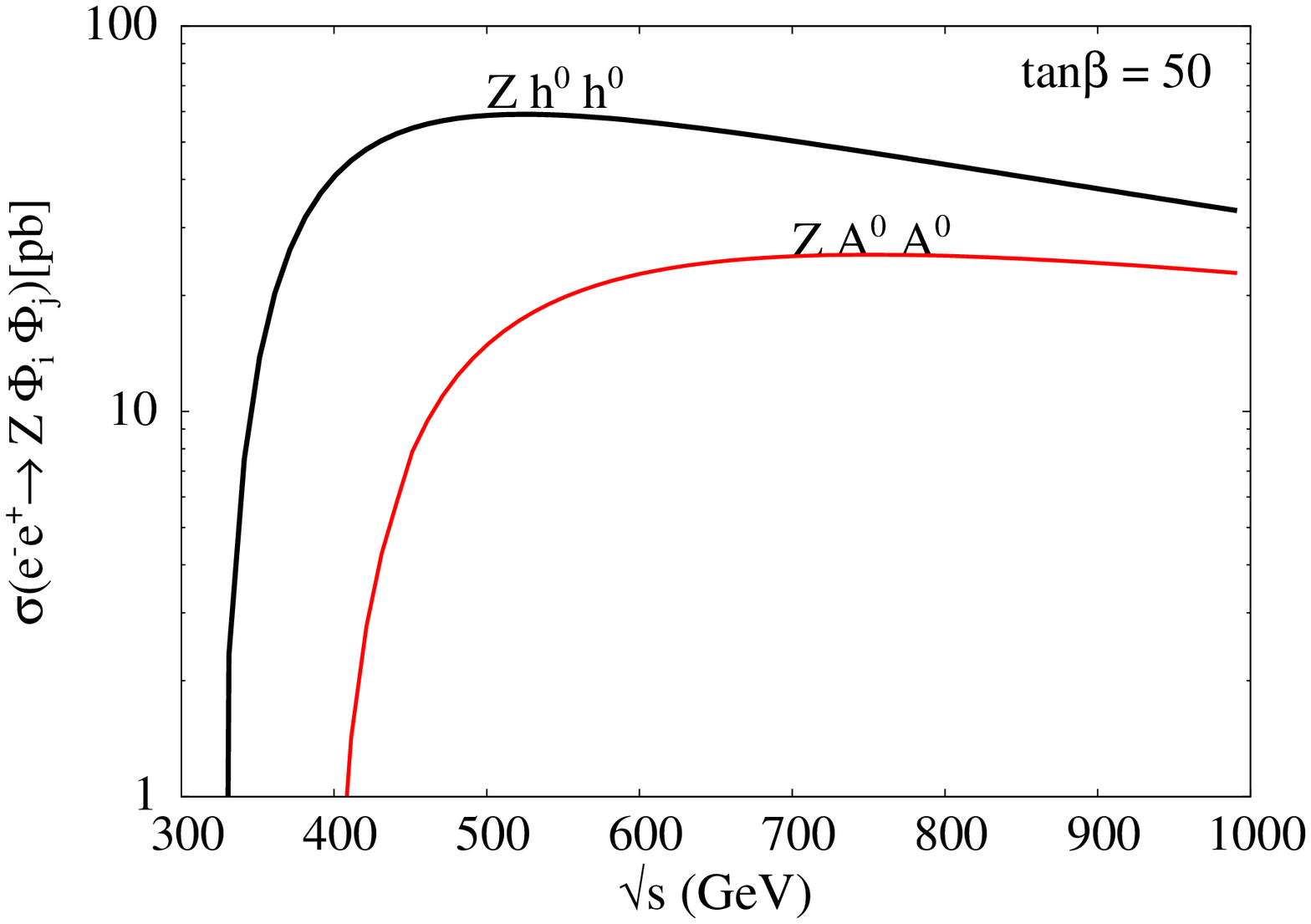}
\caption{Total tree level cross sections of $e^+ e^- \to Z \Phi_i
  \Phi_i$ in 2HDM as a function of $\sqrt{s}$ with $m_{h^0} = 120$
  GeV, $m_{H^0} = 370$ GeV, $m_{H^\pm} = 250$ GeV, $m_{A^0} = 150$
  GeV, $m_{12} = 46$ GeV, $\sin\alpha=0.6$ and 
$\tan\beta= 5$ (left) or $\tan\beta= 50$
  (right). }
\label{ffig2}
\end{figure}

In order to show the sensitivity of the cross sections to high
$\tan\beta$, we plot in Fig.~\ref{ffig2} the cases of small
$\tan\beta$ (left) and large $\tan\beta$ (right).  For small
$\tan\beta$, $e^+ e^- \to ZA^0A^0$ dominates with a few fb cross
section, $e^+ e^- \to Zh^0h^0$ and $e^+ e^- \to
Zh^0H^0$ barely reaches 0.1 fb.  This is due to the fact that the
triple Higgs couplings involved in these processes are such that
$$
(\lambda_{h^0h^0h^0}^{2HDM},
\lambda_{H^0h^0h^0}^{2HDM},
\lambda_{h^0A^0A^0}^{2HDM},
\lambda_{H^0A^0A^0}^{2HDM})
\approx (0.6,5,0.4,13) \times \lambda_{h^0h^0h^0}^{SM} ~.
$$
They are not enhanced by much except for $\lambda_{H^0A^0A^0}$ which
is $13\lambda_{h^0h^0h^0}^{SM}$ and explains the dominance of $e^+ e^-
\to ZA^0A^0$ in this case.  However, in large $\tan\beta$ cases, $e^+
e^- \to Zh^0h^0$ has the largest cross section ($\approx 60$ fb),
followed by $e^+ e^- \to ZA^0A^0$ ($\approx 20$ fb).  In this large
$\tan\beta$ case, the involved triple Higgs couplings are given by
$$
(\lambda_{h^0h^0h^0}^{2HDM},
\lambda_{H^0h^0h^0}^{2HDM},
\lambda_{h^0A^0A^0}^{2HDM},
\lambda_{H^0A^0A^0}^{2HDM})
\approx (43,-39,52,30) \times \lambda_{h^0h^0h^0}^{SM} ~.
$$
Obviously, they are much more enhanced than the low $\tan\beta$ case.

%--------------------------------------
%----------------------------------------

\begin{figure}[t!]
\centering
\includegraphics[height=2.2in]{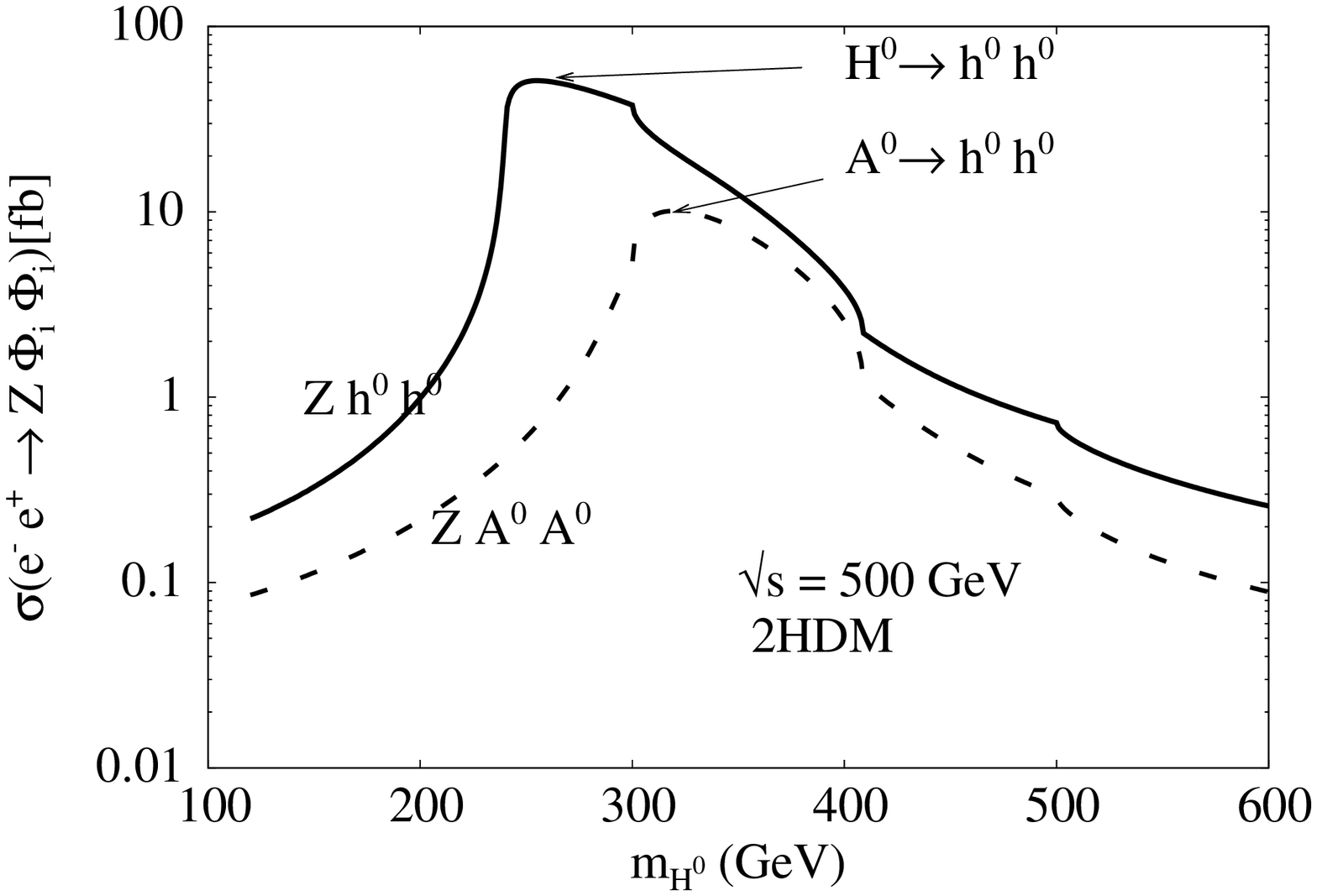}
%\hskip.5cm
\includegraphics[height=2.2in]{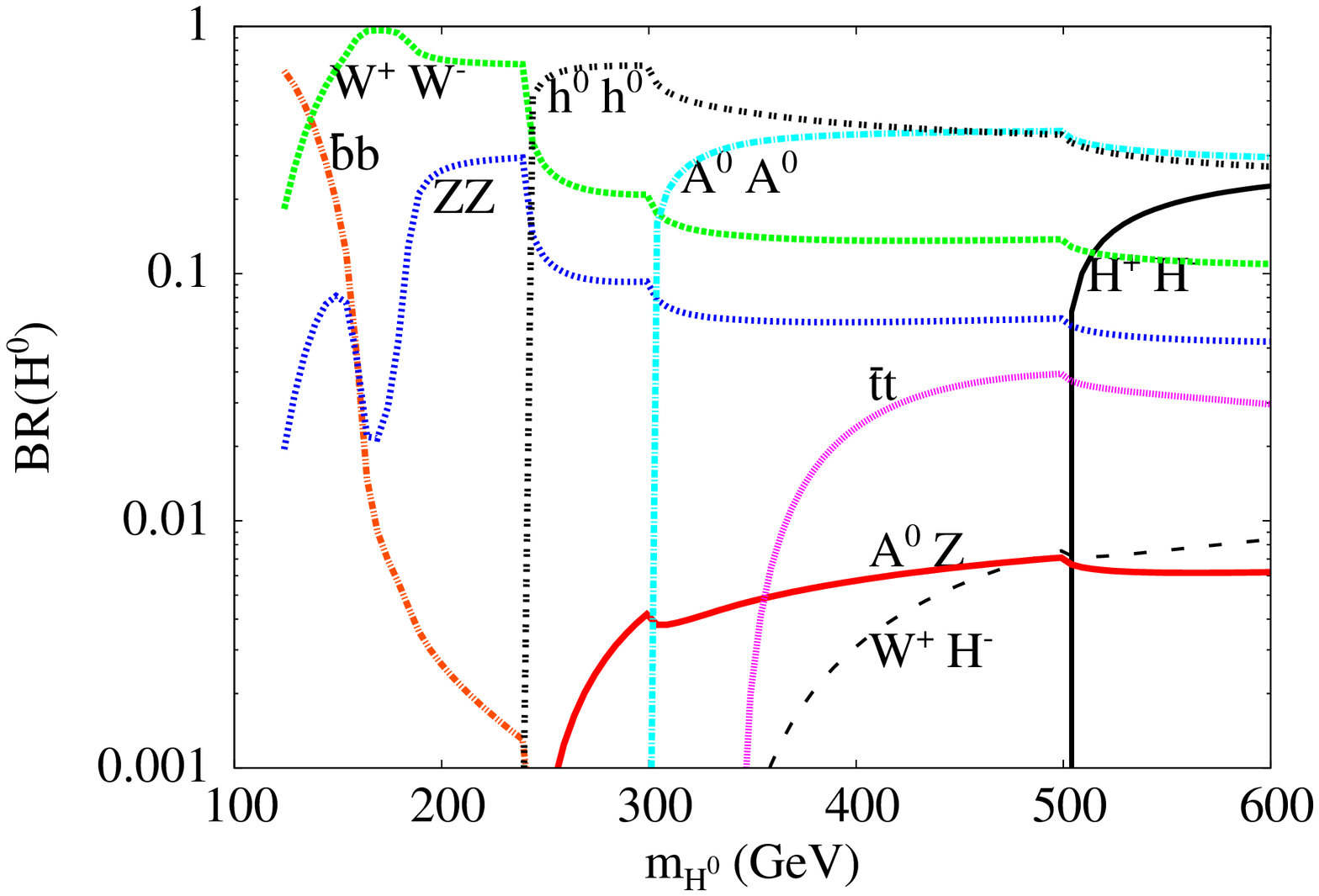}
\caption{Left plot: total cross sections for $Zh^0h^0$ and $ZA^0A^0$
  continuum and resonant production: $\sigma (e^+ e^- \to Z h^0h^0) +
  \sigma (e^+ e^- \to Z H^0) \cdot {\rm BR}(H^0\to h^0h^0,A^0A^0)$ in
  2HDM as a function of $m_{H^0}$.  Right plot: branching ratios of
  $H^0$.  Here $m_{h^0} = 120$ GeV, $m_{H^\pm} = 250$ GeV, $m_{A^0} =
  150$ GeV, $m_{12} = 0$ GeV, $\tan\beta=5$, and $\sin\alpha=0.9$.}
\label{mass1}
\end{figure}

Fig.~\ref{mass1} shows the behaviors of $e^+ e^- \to Z h^0h^0$ and
$e^+ e^- \to Z A^0A^0$ (left plot) and the branching ratios of the
CP-even $H^0$ (right plot) as a function of $m_{H^0}$ and for
$m_{h^0}$, $m_{H^\pm}$, $m_{A^0}$, $m_{12}$ = $120$, $250$, $150$, $0$
GeV, respectively, $\tan\beta=5$, $\sin\alpha=0.9$ and $\sqrt{s}=500$
GeV.  To compute the cross sections, we have summed the continuum part
$\sigma (e^+ e^- \to Z SS)$ and resonant part generated in the chain
$e^+e^-\to ZH^0\to ZSS$ via the resonant $H^0$ Higgs-strahlung, with
$S=h^0$ or $A^0$.  It is obvious that the cross sections of $e^+ e^-
\to Zh^0h^0$ and $e^+ e^- \to ZA^0A^0$ reach their maxima near the
threshold regions of $H^0\to h^0h^0$ and $H^0\to A^0A^0$.

In the right panel of Fig.~\ref{mass1}, we show the decay branching
ratios of $H^0$.  The $b\bar{b}$ mode dominates below the $WW$
threshold, while the $WW$ mode takes over for $m_{H^0} \ga 2 M_W$.
Above the $h^0h^0$ threshold, the decay $H^0\to h^0h^0$ is open and
dominates over the others at the order of 70\%.  For $m_{H^0} \ga 300$
GeV, $H^0\to A^0A^0$ is open and has also a substantial branching
ratio of ${\cal O}(20\%)$, reducing the branching ratio of $H^0\to
h^0h^0$ to the level of less than 20\%.  Note that once $H^0\to
H^+H^-$ is open, it quickly reaches the level of 15\%.

As shown in the left plot of Fig.~\ref{mass1}, there are three kinks
occurring when $m_{H^0} = 2m_{h^0}, 2m_{A^0}, 2m_{H^\pm}$ and
corresponding to the opening of the $H^0 \to h^0h^0$, $H^0 \to A^0A^0$
and $H^0 \to H^\pm H^\mp$ modes, respectively.  The cross sections
increase by about 20 fb if $h^0h^0$ and $A^0A^0$ are produced in
$e^+e^-\to ZH^0\to Zh^0h^0$ and $e^+e^-\to ZH^0\to Zh^0h^0$ via
resonant Higgs-strahlung of $H^0$.

For the double Higgs-strahlung processes $e^+e^-\to Zh^0h^0$,
$e^+e^-\to ZA^0A^0$ and $e^+e^-\to Zh^0H^0$, the dominant final states
depend on how $h^0$, $A^0$ and $H^0$ decay.  In the case where $h^0$
and $A^0$ are not heavy ($\la 125$ GeV), $h^0$ and $A^0$ will decay
to $b\bar{b}$ or $\tau^+\tau^-$, and the final states for $Zh^0h^0$
and $ZA^0A^0$ would be $Z4b$, $Z2b2\tau$ or $Z4\tau$.  In the 2HDM, it
may be possible that, if kinematically allowed, $A^0$ decays into
$Zh^0$.  In this case, the final state for $ZA^0A^0$ would be $ZA^0A^0
\to 3Zh^0h^0 \to 3Z 4b$.  For the $e^+e^-\to Zh^0H^0$ process, the
final state will be different from the previous one if the Higgs to
Higgs decays $H^0 \to h^0h^0$ and $H^0 \to A^0A^0$ are open.  As a
result, the final state would be $Zh^0h^0h^0$ and $Zh^0A^0A^0$.

We would like to stress here that the background study is beyond the scope of
this paper.  However, we point out that in the SM with small cross sections, it
has been shown in Ref.~\cite{miller} that for the $e^+e^- \to ZHH\to Z
b\bar{b}b\bar{b}$ process in the SM, the irreducible background from electroweak
and QCD processes can be suppressed down to manageable levels by using
kinematics cuts.  We expect that in the 2HDM with larger cross sections than in
SM, the signal can be easily extracted from the background as well.

%--------------------------------------
 
\subsection{Double Higgs-strahlung in the fermiophobic limit}

As commented in Section 2, in 2HDM-I the lightest CP-even Higgs boson
$h^0$ can be fermiophobic \cite{Gun,Gun1}.  This occurs when
$\alpha=\pi/2$ ($\cos\alpha=0$), so that the lightest Higgs couplings
to all fermions vanish.  In this case, the main decay channel of $h^0$
($m_{h^0}<2 M_W$) is $h^0\to \gamma \gamma$ for $m_{h^0}<M_V$ and
$h^0\to VV^*$ for $m_{h^0}>m_V$, where $V=W,Z$ \cite{Gun,brucher}.
Once the $WW$ threshold is crossed, the dominant decay mode becomes
$h^0\to W^+ W^-$.

The LEP Collaboration has already ruled out a fermiophobic Higgs with
a mass $m_{h^0} \la 104$ GeV and the $ZZh^0$ coupling similar to the
SM one \cite{LEPf}.  This constraint can be lifted if the $ZZh^0$
coupling is smaller than the SM one.

Recently, there is a study devoted to double fermiophobic Higgs bosons
production at the LHC and ILC \cite{andrewdiaz}.  For the ILC, it has
been shown that $e^+e^- \to A^0h^0$ followed by the decay $A^0\to
Zh^0$ can leads to the $Zh^0h^0$ final state that is similar to our
double Higgs-strahlung process $e^+e^- \to Zh^0h^0$.  However, the
process $A^0h^0 \to Zh^0h^0$ only depends on the gauge couplings while
the double Higgs-strahlung process $e^+e^- \to Zh^0h^0$ under consideration 
is sensitive to the 2HDM triple Higgs couplings that may enhance 
the cross section.

\begin{figure}[t!]
\centering
\includegraphics[height=2.2in]{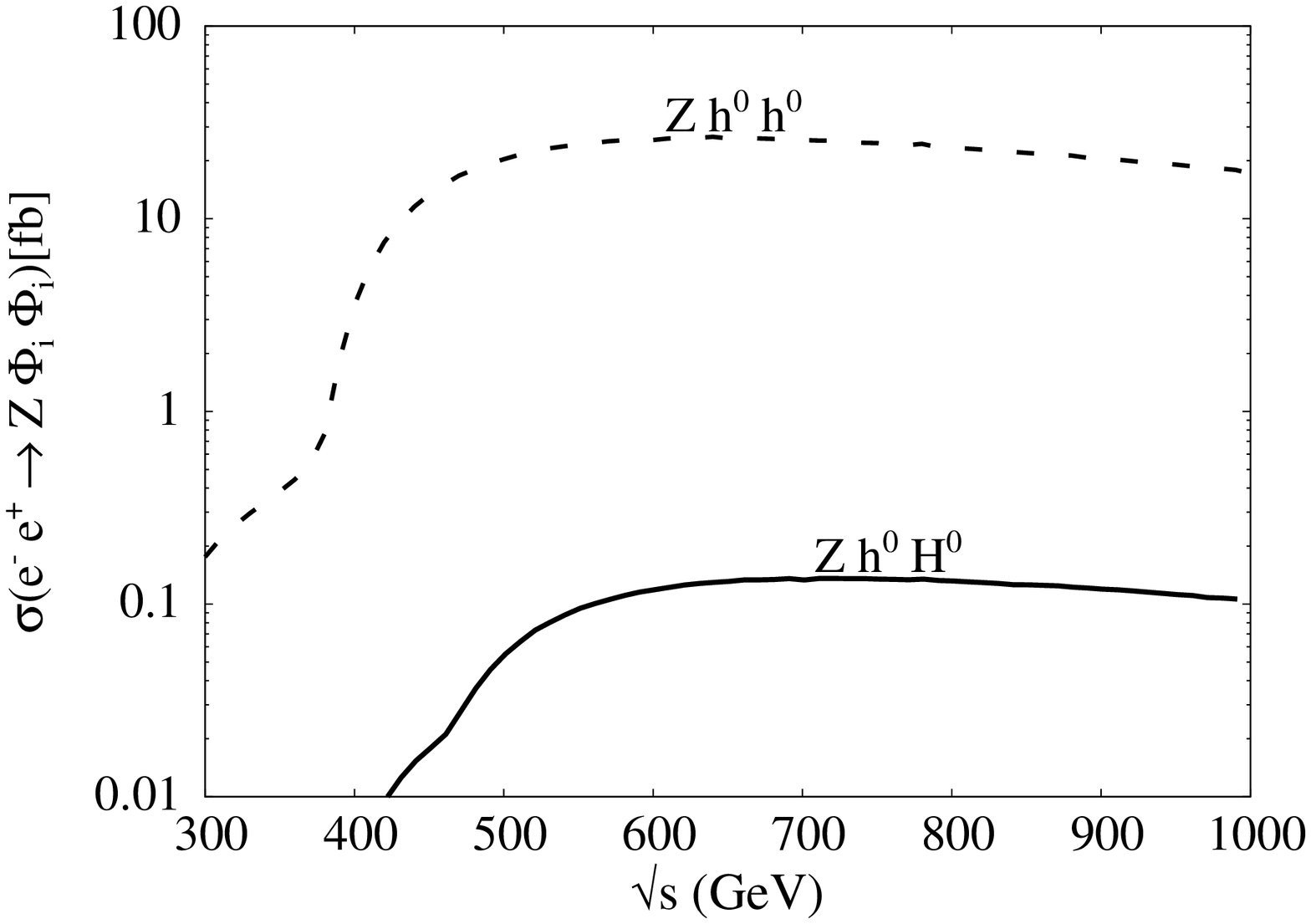}
%\hskip.5cm
\includegraphics[height=2.2in]{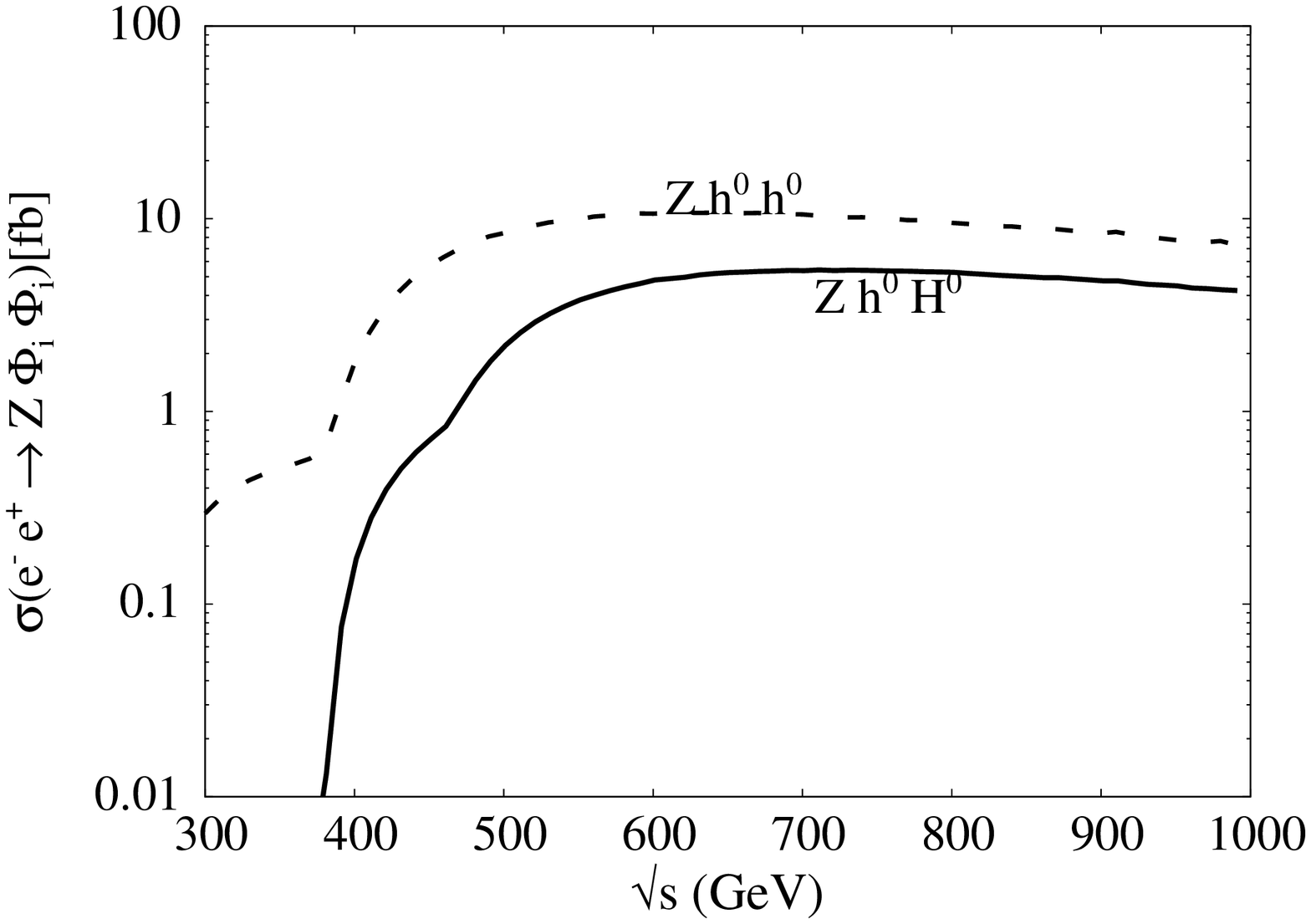}
\caption{The tree level cross sections of the $e^+ e^- \to Zh^0 h^0,
  Zh^0 H^0$ processes as a function of the center-of-mass
  energy in the fermiophobic limit with $m_{12} = 0$ GeV, $m_{h^0} =
  80$ GeV, $m_{H^0} = 160$ GeV, $m_{A^0} = 300$ GeV, $m_{H^\pm} = 250$
  GeV, and $\tan\beta = 10$ (left) or $\tan\beta = 1$ (right).}
\label{phobic}
\end{figure}

In Fig.~\ref{phobic}, we show the total cross sections of the neutral
modes $Zh^0h^0$ and $Zh^0H^0$ in the fermiophobic limit $\sin\alpha=1$
for a relatively light CP-even fermiophobic Higgs with $m_{h^0}=80$
GeV.  This possibility is not yet ruled out experimentally \cite{LEPf}
due to suppressed $ZZh^0$ coupling.  This fermiophobic limit is only
relevant for final states having at least one fermiophobic Higgs,
namely the $e^+e^-\to Zh^0h^0$ and $e^+e^-\to Zh^0H^0$ modes where the
$h^0$ may decays dominantly into two photons.  In the left panel of
Fig.~\ref{phobic} we select moderate $\tan\beta=10$, and the right
panel is for small $\tan\beta=1$.  For both channels, $Zh^0h^0$ and
$Zh^0H^0$, the cross sections are of the order of few fb and can reach
20 fb (10fb) for $Zh^0h^0$ in the case of $\tan\beta=10$
($\tan\beta=1$).  Due to phase space suppression, the cross section of
$e^+e^-\to Zh^0H^0$ is smaller than $e^+e^-\to Zh^0h^0$.

In the fermiophobic limit of 2HDM-I, one can obtain a very clear signal of
$4\gamma + X$ from the $Zh^0h^0$ mode.  For the $Zh^0H^0$ mode, the
signal depends on how $H^0$ decays.  In 2HDM, the heavy Higgs $H^0$
can decay to $W^+W^-$ or, if kinematically allowed, to $h^0h^0$.  In
the latter case, we have $ZH^0h^0 \to Zh^0h^0h^0\to Z 6\gamma$ with
$6\gamma + X$ as a distinctive signature.

To our best knowledge, there is no estimation for the $e^+ e^- \to 4 \gamma Z$
backgrounds.  We expect that such backgrounds should be small.  Moreover,
requesting four photons in our signal would be sufficient to kill the
backgrounds.

%%%%%%%%%%%%%%%%%%%%%%%%%%%%%%%%%%%%%%%
%=============================================================

\subsection{$e^+e^- \to h^0 h^0 Z$ in the decoupling limit}

In the 2HDM, the decoupling limit generally refers to the case when
all the scalar masses except one formally become infinite and the
effective theory is just the SM with one doublet (see
\cite{Gunion:2002zf} for recent discussions).  In this case, the
CP-even $h^0$ is the lightest scalar particle while the other Higgs
particles $H^0$, $A^0$ and $H^{\pm}$ are extremely heavy.  Using
purely algebraic arguments at the tree level, one can derive that the
main consequences of the decoupling limit are $\cos(\beta-\alpha)\to
0$, and the CP-even $h^0$ of the 2HDM and the SM Higgs $h_{SM}$ have
similar tree-level couplings to gauge bosons and fermions as well
\cite{Gunion:2002zf,okada}.  Obviously, the decoupling limit does not
rigorously apply to the cases where the particle masses are finite.
One can consider instead a more realistic scenario, dubbed as the
decoupling regime ~\cite{Gunion:2002zf}, where the heavy Higgs
particles have masses much larger than the $Z$ boson mass and may
escape detection in the planned experiments.

Several studies have been carried out looking for non-decoupling
effects in Higgs boson decays and Higgs self-interactions.  Large loop
effects in $h^0\to \gamma \gamma$, $h^0\to \gamma Z$ and $h^0\to
b\bar{b}$ have been pointed out for the 2HDM~\cite{maria,indirect1}
and may provide indirect information on the Higgs masses and the
involved triple Higgs couplings such as $\lambda_{h^0H^+H^-}$,
$\lambda_{h^0H^0H^0}$, $\lambda_{h^0A^0A^0}$ and
$\lambda_{h^0h^0h^0}$.  The non-decoupling contributions to the triple
Higgs self-couplings $\lambda_{h^0h^0h^0}$ have been investigated in
the 2HDM in Ref.~\cite{okada}, revealing large non-decoupling effects.

In this section, we will show that the large non-decoupling effects in
$\lambda_{h^0h^0h^0}$ modify the double Higgs-strahlung $e^+ e^- \to
Zh^0h^0$ cross section and make it larger than the SM expectation.  We
will focus on the scenario where all the Higgs particles of the 2HDM,
except for the lightest CP-even Higgs, are heavy and can escape from
detection at the first stage of next generation colliders.

%-------------------------------------------------------------------
\begin{figure}[t!]
\centering
\includegraphics[height=2.2in]{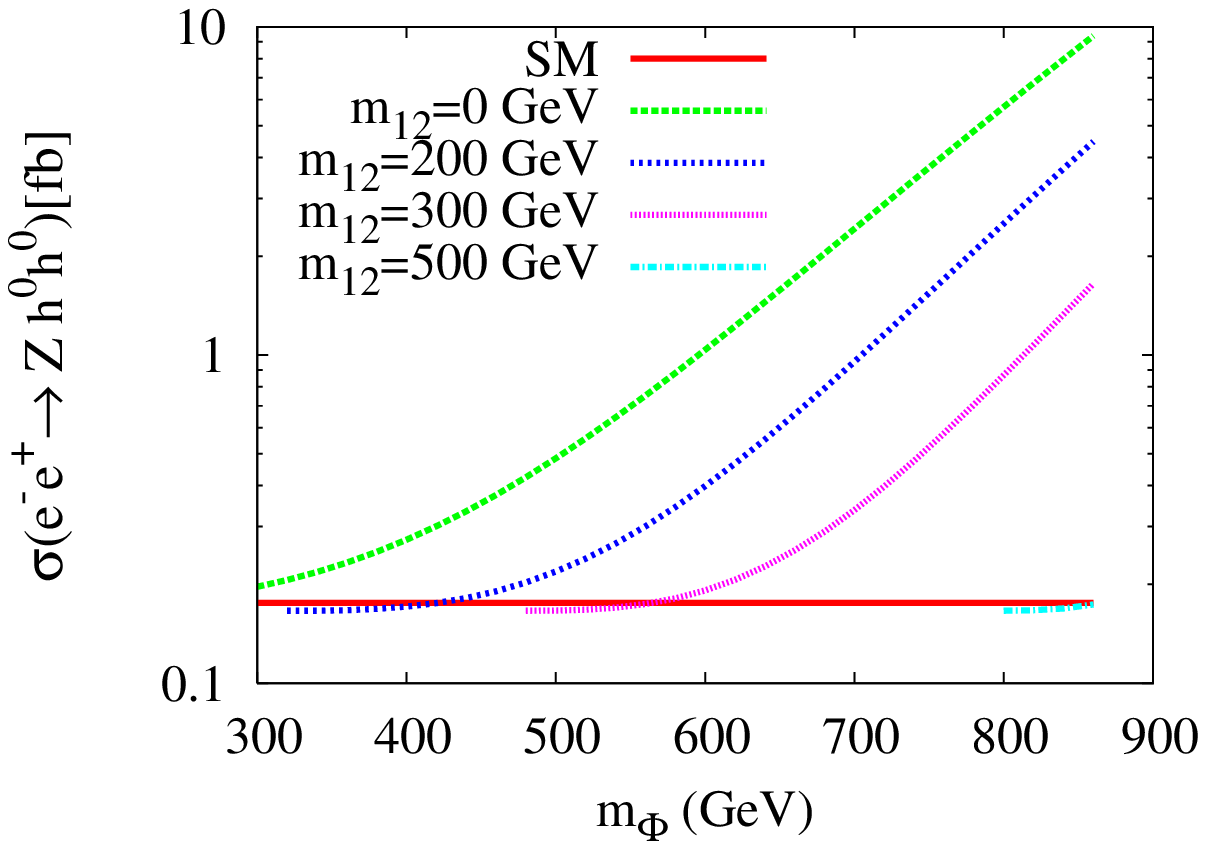}
%\hskip.5cm
\includegraphics[height=2.2in]{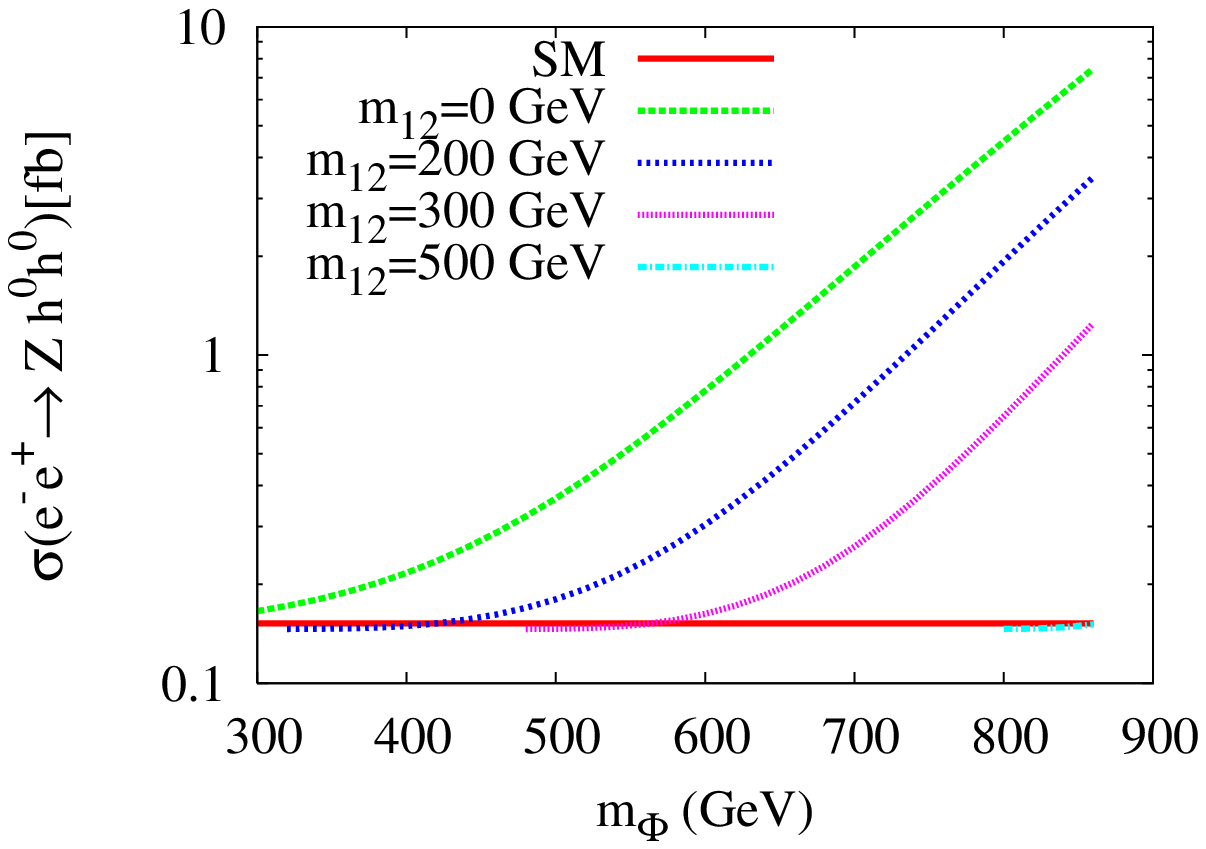}
\caption {$\sigma(e^+ e^- \to Zh^0h^0)$ in units of fb as a function
  of $M_\Phi$ at $\sqrt{s}=500$ GeV (left) and $\sqrt{s}=800$ GeV
  (right) for different values of $m_{12}$.  For both plots,
  $m_{h^0}=120$ GeV, $\tan\beta=2$ and $\cos(\beta-\alpha)=0$.}
\label{plot4}
\end{figure}
%-----------

It is easy to check that in the decoupling limit $\beta-\alpha \to
\pi/2$, the triple Higgs coupling $\lambda_{h^0h^0h^0}$ given in
Eq.~(\ref{lll}) reduces to its SM value $\lambda_{h^0h^0h^0}^{SM}=- 3
m_{h^0}^2 / v$.  In Ref.~\cite{okada}, using the Feynman diagrammatic
method, it has been demonstrated that the one-loop leading
contributions originated from the heavy Higgs boson loops and the top
quark loops to the effective $\lambda_{h^0h^0h^0}$ coupling can be
written as
\begin{eqnarray}
 \lambda_{h^0h^0h^0}^{eff}({\rm 2HDM}) \!\!&=&\!\! \frac{3 m_{h^0}^2}{v}
      \left\{ 1  
     + \frac{m_{H^0}^4}{12 \pi^2 m_{h^0}^2 v^2} 
     \left(1 - \frac{M^2}{m_{h^0}^2}\right)^3 
     + \frac{m_{A^0}^4}{12 \pi^2 m_{h^0}^2 v^2} 
     \left(1 - \frac{M^2}{m_{A^0}^2}\right)^3 \right.\nonumber\\
&&\left.
\!\!\!\!\!\!\!\!\!\!\!\!\!\!\!\!\!\!\!\!\!\!\!\!\!\!\!\!\!\!
\!\!\!\!\!\!\!\!\!\!\!\!\!\!\!\!\!\!
    + \frac{m_{H^\pm}^4}{6 \pi^2 m_{h^0}^2 v^2} 
    \left(1 - \frac{M^2}{m_{H^\pm}^2}\right)^3
     - \frac{N_c m_t^4}{3 \pi^2 m_{h^0}^2 v^2} + 
      {\cal O} \left(\frac{p^2_i m_\Phi^2}{m_{h^0}^2 v^2},
     \;\frac{m_\Phi^2}{v^2},
     \;\frac{p^2_i m_t^2}{m_{h^0}^2 v^2},  
     \;\frac{m_t^2}{v^2}  \right)
      \right\} ~,
\label{ceff}    
\end{eqnarray}
where $M=m_{12}/\sqrt{\sin\be\cos\be}$, $m_\Phi^{}$ and $p_i$
represent the mass of $H^0$, $A^0$ or $H^\pm$ and the momenta of external
Higgs lines, respectively, $N_c$ denotes the number of colors, and
$m_t$ is the mass of top quark.  We note that in Eq.~(\ref{ceff})
$m_{h^0}$ is the renormalized physical mass of the lightest CP-even
Higgs boson $h^0$.

At the tree level, one can show that in the decoupling limit the
$ZZH^0$ coupling approaches 0 and the $e^+e^- \to Zh^0h^0$ amplitude
reduces exactly to the SM result.  For our calculation of the cross
section of $e^+e^- \to Zh^0h^0$ in the decoupling limit, we ignore
one-loop corrections to the $ZZh^0$ coupling and replace the
$h^0h^0h^0$ coupling by its effective coupling given in
Eq.~(\ref{ceff}).  In fact, it has been shown in Ref.~\cite{okada}
that the non-decoupling effect in $ZZh^0$ is at a few percent level in
the case of $\cos(\beta-\alpha)=0$.

In Fig.~\ref{plot4}, we show the cross section of $e^+e^- \to Zh^0h^0$
in the SM and 2HDM in its decoupling regime as a function of
$m_{\Phi}=m_{A^0}=m_{H^0}=m_{H^\pm}$.  In the calculation of the SM
cross section, we take into account the one-loop leading contributions
originated from top quarks given by \cite{okada}
\begin{eqnarray}
 \lambda_{hhh}^{eff}(SM) \!\!&=&\!\! \frac{3 m_{h^0}^2}{v}
      \left\{ 1 - \frac{N_c m_t^4}{3 \pi^2 m_{h^0}^2 v^2} + 
      {\cal O} \left( \frac{m_t^2}{v^2}  \right)
      \right\}, \label{smeff}    
\end{eqnarray}
For our choice of $m_{h^0}=120$ GeV, the SM cross section is tiny,
$\approx 0.2$ fb and $0.16$ fb for $\sqrt{s}=500, 800$ GeV,
respectively.  As is clear in Fig.~\ref{plot4}, the 2HDM contributions
can enhance the cross section by more than one order of magnitude and
reach a few fb for small $m_{12}$ and large $m_{\Phi}$.  As shown in
both panels of Fig.~\ref{plot4}, we get maximum non-decoupling effect
for $m_{12}=0$.  It is further amplified with large $m_{\Phi}$.  The
plots are cut due to the perturbativity and vacuum stability
constraints.  The perturbativity constraints on $\lambda_i$ do not
allow $m_\Phi$ larger than $850$ GeV.  In the case of $m_{12}=500$
GeV, the vacuum stability condition does not allow $m_\Phi$ to be less
than 800 GeV.  The sensitivity of cross section to $\tan\beta$ is mild,
since $\tan\beta$ only enters the $h^0h^0h^0$ coupling through
$M^2=m_{12}^2/(\sin\beta\cos\beta)$.  Moreover, $\tan\beta$ is
constrained by perturbativity to be rather moderate.

\section{Summary}
\label{sec:summary}

We have studied the triple Higgs couplings $\lambda_{\Phi_i \Phi_j
  \Phi_k}$ and the double Higgs-strahlung production $e^+e^- \to Z
\Phi_i \Phi_j$ at linear collider in the framework of general Two
Higgs Doublet Models.  We have quantified the sizes of the triple
Higgs couplings compared to the SM triple Higgs coupling.  We also
show that after taking into account the perturbativity and vacuum
stability constraints on the 2HDM parameters, it is possible to
enhance the triple Higgs coupling $\lambda_{h^0h^0h^0}$ up to 15 times
or more than the corresponding SM coupling.  If the Higgs bosons
$\Phi=$ $h^0$, $H^0$, $A^0$ and $H^\pm$ are not too heavy, the double
Higgs-strahlung cross sections $e^+e^- \to Z \Phi_i \Phi_j$ can be
substantial, at the level of a few hundred fb, and provide some
information on the triple Higgs couplings.

We have also studied the double Higgs-strahlung processes $e^+e^-\to
Zh^0h^0$ and $e^+e^-\to ZH^0h^0$ in the fermiophobic limit of 2HDM-I 
where the collider signature can be very distinctive with the final state
$Z2\gamma$, $Z4\gamma$, or $Z6\gamma$ if $H^0\to h^0h^0$.  We also
analyze the double Higgs-strahlung process $e^+e^- \to Z h^0h^0$ in
the decoupling limit where $h^0$ mimics the SM Higgs boson.  It is
shown that in this limit, the cross section can be enhanced by about
two orders of magnitude, which is much larger than the MSSM
enhancement.  Observations of such large cross sections would
definitely indicate that the Higgs sector is 2HDM-like.

%--------------------------------------------
\section{acknowledgments}

We would like to thank A.G. Akeroyd, S. Kanemura, R.R. Santos and
Eibun Senaha for valuable discussions and comments.  AA, RB and CWC
are supported by the National Science Council of R.O.C. under Grant
Nos. NSC 96-2811-M-008-020, NSC 96-2811-M-033-005, and NSC
96-2112-M-008-001, respectively.  CWC also thanks the Institute of
Physics at National Chiao-Tung Univ. for the hospitality during his
visit.

%======================================================

%--------------------------------------

\begin{thebibliography}{2006}


\bibitem{Higgs:1964pj}
  P.~W.~Higgs,
  %``BROKEN SYMMETRIES AND THE MASSES OF GAUGE BOSONS,''
  Phys.\ Rev.\ Lett.\  {\bf 13} (1964) 508.
  %%CITATION = PRLTA,13,508;%%
 G.~S.~Guralnik, C.~R.~Hagen and T.~W.~B.~Kibble,
  %``GLOBAL CONSERVATION LAWS AND MASSLESS PARTICLES,''
  Phys.\ Rev.\ Lett.\  {\bf 13}, 585 (1964).
  %%CITATION = PRLTA,13,585;%%
 F.~Englert and R.~Brout,
  %``BROKEN SYMMETRY AND THE MASS OF GAUGE VECTOR MESONS,''
  Phys.\ Rev.\ Lett.\  {\bf 13} (1964) 321.
  %%CITATION = PRLTA,13,321;%%

 
\bibitem{Ruwiedel:2007tv}
M.~Duhrssen, S.~Heinemeyer, H.~Logan, D.~Rainwater, 
G.~Weiglein and D.~Zeppenfeld,
  %``Extracting Higgs boson couplings from LHC data,''
  Phys.\ Rev.\  D {\bf 70} (2004) 113009
  [arXiv:hep-ph/0406323];
  %%CITATION = PHRVA,D70,113009;%%
  C.~Ruwiedel and f.~t.~A.~Collaborations,
  %``Prospects for the Determination of Higgs Boson Properties at the LHC,''
  arXiv:0710.1954 [hep-ph].
  %%CITATION = ARXIV:0710.1954;%%

\bibitem{ILCLHC}  
G.~Weiglein {\it et al.}  [LHC/LC Study Group],
  %``Physics interplay of the LHC and the ILC,''
  Phys.\ Rept.\  {\bf 426} (2006) 47
  [arXiv:hep-ph/0410364].
  %%CITATION = PRPLC,426,47;%%


\bibitem{Gun} 
  J.~F.~Gunion, H.~E.~Haber, G.~L.~Kane and S.~Dawson,
  ``THE HIGGS HUNTER'S GUIDE,''(Addison--Wesley, Reading, 1990).
  %%CITATION = BNL-41644;%%




\bibitem{abdel1}
A.~Djouadi,
  %``The anatomy of electro-weak symmetry breaking. I: The Higgs boson in  the
  %standard model,''
  arXiv:hep-ph/0503172.
  %%CITATION = HEP-PH/0503172;%%

\bibitem{abdel2}
A.~Djouadi,
  %``The anatomy of electro-weak symmetry breaking. II: 
  %The Higgs bosons in  the
  %minimal supersymmetric model,''
  arXiv:hep-ph/0503173.
  %%CITATION = HEP-PH/0503173;%%

\bibitem{MSSMHIGGS}
A.~Djouadi, P.~M.~Zerwas and H.~E.~Haber,
  %``Multiple production of MSSM neutral Higgs bosons at high-energy  e+ e-
  %colliders,''
  arXiv:hep-ph/9605437.
  %%CITATION = HEP-PH/9605437;%%

\bibitem{tripleSM}
 A.~Djouadi, W.~Kilian, M.~Muhlleitner and P.~M.~Zerwas,
  %``Testing Higgs Self-couplings at e^+e^- Linear Colliders,SM and MSSM''
  Eur.\ Phys.\ J.\  C {\bf 10}, 27 (1999)
  [arXiv:hep-ph/9903229].
  %%CITATION = EPHJA,C10,27;%%

\bibitem{Osland:1998hv}
  P.~Osland and P.~N.~Pandita,
  %``Measuring the trilinear couplings of MSSM neutral Higgs bosons at
  %high-energy e+ e- colliders,''
  Phys.\ Rev.\  D {\bf 59}, 055013 (1999)
  [arXiv:hep-ph/9806351].
  %%CITATION = PHRVA,D59,055013;%%
F.~Boudjema and A.~Semenov,
  %``Measurements of the SUSY Higgs self-couplings and the reconstruction of
  %the Higgs potential,''
  Phys.\ Rev.\  D {\bf 66}, 095007 (2002)
  [arXiv:hep-ph/0201219].
  %%CITATION = PHRVA,D66,095007;%%

\bibitem{miller}
D.~J.~Miller and S.~Moretti,
  %``Can the trilinear Higgs self-coupling be measured at future linear
  %colliders?,''
  Eur.\ Phys.\ J.\  C {\bf 13} (2000) 459
  [arXiv:hep-ph/9906395].
  %%CITATION = EPHJA,C13,459;%%

\bibitem{tripleMSSMLHC}
A.~Djouadi, W.~Kilian, M.~Muhlleitner and P.~M.~Zerwas,
  %``Production of neutral Higgs-boson pairs at LHC,''
  Eur.\ Phys.\ J.\  C {\bf 10} (1999) 45
  [arXiv:hep-ph/9904287].
  %%CITATION = EPHJA,C10,45;%%
T.~Plehn, M.~Spira and P.~M.~Zerwas,
  %``Pair Production of Neutral Higgs Particles in Gluon--Gluon Collisions,''
  Nucl.\ Phys.\  B {\bf 479}, 46 (1996)
  [Erratum-ibid.\  B {\bf 531}, 655 (1998)]
  [arXiv:hep-ph/9603205];
  %%CITATION = NUPHA,B479,46;%%
%\bibitem{Dai:1995cb}
  J.~Dai, J.~F.~Gunion and R.~Vega,
  %``Detection of the Minimal Supersymmetric Model Higgs Boson $H~0$ in its
  %$h~0h~0\to 4b$ and $A~0A~0\to 4b$ Decay Channels,''
  Phys.\ Lett.\  B {\bf 371}, 71 (1996)
  [arXiv:hep-ph/9511319] and
  %%CITATION = PHLTA,B371,71;%%
%  J.~Dai, J.~F.~Gunion and R.~Vega,
  %``Detection of neutral MSSM Higgs bosons in four-b final states at the
  %Tevatron and the LHC: An update,''
  Phys.\ Lett.\  B {\bf 387}, 801 (1996)
  [arXiv:hep-ph/9607379];
  %%CITATION = PHLTA,B387,801;%%
%\bibitem{Lafaye:2000ec}
  R.~Lafaye, D.~J.~Miller, M.~Muhlleitner and S.~Moretti,
  %``Double Higgs production at TeV colliders in the minimal supersymmetric
  %standard model,''
  arXiv:hep-ph/0002238;
  %%CITATION = HEP-PH/0002238;%%
%\bibitem{BarrientosBendezu:1999gp}
  A.~A.~Barrientos Bendezu and B.~A.~Kniehl,
  %``H+ H- pair production at the Large Hadron Collider,''
  Nucl.\ Phys.\  B {\bf 568}, 305 (2000)
  [arXiv:hep-ph/9908385] and
  %%CITATION = NUPHA,B568,305;%%
%\bibitem{BarrientosBendezu:2001di}
%  A.~A.~Barrientos Bendezu and B.~A.~Kniehl,
  %``Pair production of neutral Higgs bosons at the CERN Large Hadron
  %Collider,''
  Phys.\ Rev.\  D {\bf 64}, 035006 (2001)
  [arXiv:hep-ph/0103018];
  %%CITATION = PHRVA,D64,035006;%%
%\bibitem{Baur:2002rb}
  U.~Baur, T.~Plehn and D.~L.~Rainwater,
  %``Measuring the Higgs boson self coupling at the LHC and finite top mass
  %matrix elements,''
  Phys.\ Rev.\ Lett.\  {\bf 89}, 151801 (2002)
  [arXiv:hep-ph/0206024],
  %%CITATION = PRLTA,89,151801;%%
%\bibitem{Baur:2002qd}
%  U.~Baur, T.~Plehn and D.~L.~Rainwater,
  %``Determining the Higgs boson self coupling at hadron colliders,''
  Phys.\ Rev.\  D {\bf 67}, 033003 (2003)
  [arXiv:hep-ph/0211224],
  %%CITATION = PHRVA,D67,033003;%%
%\bibitem{Baur:2003gpa}
%  U.~Baur, T.~Plehn and D.~L.~Rainwater,
  %``Examining the Higgs boson potential at lepton and hadron colliders: A
  %comparative analysis,''
  Phys.\ Rev.\  D {\bf 68}, 033001 (2003)
  [arXiv:hep-ph/0304015], and 
  %%CITATION = PHRVA,D68,033001;%%
%\bibitem{Baur:2003gp}
%  U.~Baur, T.~Plehn and D.~L.~Rainwater,
  %``Probing the Higgs self-coupling at hadron colliders using rare decays,''
  Phys.\ Rev.\  D {\bf 69}, 053004 (2004)
  [arXiv:hep-ph/0310056];
  %%CITATION = PHRVA,D69,053004;%%
%\bibitem{Moretti:2003px}
%  S.~Moretti and J.~Rathsman,
  %``Pair production of charged Higgs bosons in association with bottom  quark
  %pairs at the Large Hadron Collider,''
  Eur.\ Phys.\ J.\  C {\bf 33}, 41 (2004)
  [arXiv:hep-ph/0308215];
  %%CITATION = EPHJA,C33,41;%%
%\bibitem{Moretti:2001pp}
  S.~Moretti,
  %``Pair production of charged Higgs scalars from electroweak gauge boson
  %fusion,''
  J.\ Phys.\ G {\bf 28}, 2567 (2002)
  [arXiv:hep-ph/0102116];
  %%CITATION = JPHGB,G28,2567;%%
%\bibitem{Moretti:2007ca}
  M.~Moretti, S.~Moretti, F.~Piccinini, R.~Pittau and J.~Rathsman,
  %``Vector-Boson Production of Light Higgs Pairs in 2-Higgs Doublet Models,''
  JHEP {\bf 0712}, 075 (2007)
  [arXiv:0706.4117 [hep-ph]];
  %%CITATION = JHEPA,0712,075;%%
%\bibitem{Moretti:2004wa}
  M.~Moretti, S.~Moretti, F.~Piccinini, R.~Pittau and A.~D.~Polosa,
  %``Higgs boson self-couplings at the LHC as a probe of extended Higgs
  %sectors,''
  JHEP {\bf 0502}, 024 (2005)
  [arXiv:hep-ph/0410334].
  %%CITATION = JHEPA,0502,024;%%

\bibitem{Dubinin:1998nt}
  M.~N.~Dubinin and A.~V.~Semenov,
  %``Triple and quartic interactions of Higgs bosons in the general
  %two-Higgs-doublet model,''
  arXiv:hep-ph/9812246.
  %%CITATION = HEP-PH/9812246;%%

\bibitem{Dubinin:2002nx}
  M.~N.~Dubinin and A.~V.~Semenov,
  %``Triple and quartic interactions of Higgs bosons in the  two-Higgs-doublet
  %model with CP-violation,''
  Eur.\ Phys.\ J.\  C {\bf 28}, 223 (2003)
  [arXiv:hep-ph/0206205];
  %%CITATION = EPHJA,C28,223;%%
%\bibitem{Osland:2008aw}
  P.~Osland, P.~N.~Pandita and L.~Selbuz,
  %``Trilinear Higgs couplings in the two Higgs doublet model with CP
  %violation,''
  arXiv:0802.0060 [hep-ph].
  %%CITATION = ARXIV:0802.0060;%%

\bibitem{sola}
 G.~Ferrera, J.~Guasch, D.~Lopez-Val and J.~Sola,
  %``Triple Higgs boson production in the Linear Collider,''
  arXiv:0707.3162 [hep-ph].
  %%CITATION = ARXIV:0707.3162;%%

\bibitem{Gunion:2002zf}
  J.~F.~Gunion and H.~E.~Haber,
%``The CP-conserving two-Higgs-doublet model: The approach to the  
%decoupling limit,''
  Phys.\ Rev.\  D {\bf 67}, 075019 (2003)
  [arXiv:hep-ph/0207010].
  %%CITATION = PHRVA,D67,075019;%%

\bibitem{am}
 A.~Arhrib and G.~Moultaka,
  %``Radiative corrections to e+ e- --> H+ H-: THDM versus MSSM,''
  Nucl.\ Phys.\  B {\bf 558}, 3 (1999)
  [arXiv:hep-ph/9808317].
  %%CITATION = NUPHA,B558,3;%%
 
\bibitem{aan}
A.~G.~Akeroyd, A.~Arhrib and E.~Naimi,
  %``Radiative corrections to the decay H+ --> W+ A0,''
  Eur.\ Phys.\ J.\  C {\bf 20}, 51 (2001)
  [arXiv:hep-ph/0002288].
  %%CITATION = EPHJA,C20,51;%%

\bibitem{santos} 
 L.~Brucher and R.~Santos,
  %``Experimental signatures of fermiophobic Higgs bosons,''
  Eur.\ Phys.\ J.\  C {\bf 12} (2000) 87
  [arXiv:hep-ph/9907434].
  %%CITATION = EPHJA,C12,87;%%

\bibitem{Glashow:1976nt}
  S.~L.~Glashow and S.~Weinberg,
  %``Natural Conservation Laws For Neutral Currents,''
  Phys.\ Rev.\  D {\bf 15} (1977) 1958.
  %%CITATION = PHRVA,D15,1958;%%

\bibitem{rodrigues} 
 T.~P.~Cheng and M.~Sher,
  %``Mass Matrix Ansatz and Flavor Nonconservation in Models 
%with Multiple Higgs Doublets,''
  Phys.\ Rev.\  D {\bf 35} (1987) 3484.
  %%CITATION = PHRVA,D35,3484;%%
 D.~Atwood, L.~Reina and A.~Soni,
  %``Phenomenology of two Higgs doublet models with flavor changing neutral
  %currents,''
  Phys.\ Rev.\  D {\bf 55} (1997) 3156;
  %%CITATION = PHRVA,D55,3156;%%
R.~Diaz, R.~Martinez and J.~A.~Rodriguez,
  %``Lepton flavor violation in the two Higgs doublet model type III,''
  Phys.\ Rev.\  D {\bf 63} (2001) 095007;
  %%CITATION = PHRVA,D63,095007;%%
A.~E.~Carcamo, R.~Martinez and J.~A.~Rodriguez,
  %``Different kind of textures of Yukawa coupling matrices in the two Higgs
  %doublet model type III,''
  Eur.\ Phys.\ J.\  C {\bf 50} (2007) 935.
  %%CITATION = EPHJA,C50,935;%%

\bibitem{Gun1} 
Thomas J. Weiler, talk given at 8th Vanderbilt Int. Conf. on High 
Energy Physics, Nashville, Oct 8-10, 1987;
Published in Nashville Conf.1987:0219 
%%CITATION = C87/10/08;%%
H.~E.~Haber, G.~L.~Kane and T.~Sterling,
%``The Fermion Mass Scale And Possible Effects Of Higgs Bosons On Experimental
%Observables,''
Nucl.\ Phys.\  B {\bf 161}, 493 (1979).
%%CITATION = NUPHA,B161,493;%%


\bibitem{Rhoparam}
A.~Denner, R.~J.~Guth, W.~Hollik and J.~H.~Kuhn,
  %``The Z width in the two Higgs doublet model,''
  Z.\ Phys.\  C {\bf 51}, 695 (1991).
  %%CITATION = ZEPYA,C51,695;%%

\bibitem{pdg4}
S.~Eidelman {\it et al.}  [Particle Data Group],
%``Review of particle physics,''
  Phys.\ Lett.\ B {\bf 592} (2004) 1.
%%CITATION = PHLTA,B592,1;%%
%and 2005 update at http://pdg.lbl.gov/.

\bibitem{berger} 
V.~D.~Barger, J.~L.~Hewett and R.~J.~N.~Phillips,
  %``NEW CONSTRAINTS ON THE CHARGED HIGGS SECTOR IN TWO HIGGS DOUBLET MODELS,''
  Phys.\ Rev.\  D {\bf 41} (1990) 3421.
  %%CITATION = PHRVA,D41,3421;%%


\bibitem{Oslandk}  A.~W.~El Kaffas, O.~M.~Ogreid and P.~Osland,
  %``Profile of Two-Higgs-Doublet-Model Parameter Space,''
  arXiv:0709.4203 [hep-ph].
  %%CITATION = ARXIV:0709.4203;%%
A.~Wahab El Kaffas, P.~Osland and O.~Magne Ogreid,
  %``Constraining the Two-Higgs-Doublet-Model parameter space,''
  Phys.\ Rev.\  D {\bf 76}, 095001 (2007)
  [arXiv:0706.2997 [hep-ph]].
  %%CITATION = PHRVA,D76,095001;%%

\bibitem{unit1}
S.~Kanemura, T.~Kubota and E.~Takasugi,
%``Lee-Quigg-Thacker bounds for Higgs boson masses in a two doublet model,''
Phys.\ Lett.\  B {\bf 313}, 155 (1993)
[arXiv:hep-ph/9303263].
%%CITATION = PHLTA,B313,155;%%

\bibitem{abdesunit}  
A.~G.~Akeroyd, A.~Arhrib and E.~M.~Naimi,
  %``Note on tree-level unitarity in the general two Higgs doublet model,''
  Phys.\ Lett.\  B {\bf 490}, 119 (2000)
  [arXiv:hep-ph/0006035].
  %%CITATION = PHLTA,B490,119;%%
A.~Arhrib,
  %``Unitarity constraints on scalar parameters of the standard and two  Higgs
  %doublets model,''
  arXiv:hep-ph/0012353.
  %%CITATION = HEP-PH/0012353;%%
 J.~Horejsi and M.~Kladiva,
  %``Tree-unitarity bounds for THDM Higgs masses revisited,''
  Eur.\ Phys.\ J.\  C {\bf 46}, 81 (2006)
  [arXiv:hep-ph/0510154].
  %%CITATION = EPHJA,C46,81;%%


\bibitem{vac1} 
M.~Sher,
  %``Electroweak Higgs Potentials And Vacuum Stability,''
  Phys.\ Rept.\  {\bf 179}, 273 (1989).
  %%CITATION = PRPLC,179,273;%%
S.~Kanemura, T.~Kasai and Y.~Okada,
  %``Mass bounds of the lightest CP-even Higgs boson in the  two-Higgs-doublet
  %model,''
  Phys.\ Lett.\ B {\bf 471}, 182 (1999)
  [arXiv:hep-ph/9903289].
  %%CITATION = HEP-PH 9903289;%%

\bibitem{vac2} 
A.~Barroso, P.~M.~Ferreira and R.~Santos,
  %``Some remarks on tree-level vacuum stability in two Higgs doublet models,''
  arXiv:hep-ph/0507329.
  %%CITATION = HEP-PH 0507329;%%
P.~M.~Ferreira, R.~Santos and A.~Barroso,
  %``Stability of the tree-level vacuum in two Higgs doublet models against
  %charge or CP spontaneous violation,''
  Phys.\ Lett.\ B {\bf 603}, 219 (2004)
  [arXiv:hep-ph/0406231].
  %%CITATION = HEP-PH 0406231;%%


\bibitem{Abdallah:2003wd}
  J.~Abdallah {\it et al.}  [DELPHI Collaboration],
%``Search for charged Higgs bosons at LEP in general two Higgs doublet
%models,''
  Eur.\ Phys.\ J.\  C {\bf 34} (2004) 399; OPAL Phys. Note PN445 (2000);
OPAL Phys. Note PN472 (2001). 
%[arXiv:hep-ex/0404012].
%%CITATION = EPHJA,C34,399;%%

\bibitem{opal}
G.~Abbiendi {\it et al.}  [OPAL Collaboration],
  %``Flavour independent h0 A0 search and two Higgs doublet model
  %interpretation of neutral Higgs boson searches at LEP,''
  Eur.\ Phys.\ J.\  C {\bf 40} (2005) 317
  [arXiv:hep-ex/0408097].
  %%CITATION = EPHJA,C40,317;%%


\bibitem{delphi}
 J.~Abdallah {\it et al.}  [DELPHI Collaboration],
  %``Searches for neutral Higgs bosons in extended models,''
  Eur.\ Phys.\ J.\  C {\bf 38}, 1 (2004)
  [arXiv:hep-ex/0410017].
  %%CITATION = EPHJA,C38,1;%%


\bibitem{sola1}
S.~Bejar, J.~Guasch and J.~Sola,
  %``Higgs boson flavor-changing neutral decays into top quark in a general
  %two-Higgs-doublet model,''
  Nucl.\ Phys.\  B {\bf 675} (2003) 270
  [arXiv:hep-ph/0307144].
  %%CITATION = NUPHA,B675,270;%% 


\bibitem{feynarts}
T.~Hahn, Comput.\ Phys.\ Commun.\  {\bf 140}, 418 (2001);
%%CITATION = HEP-PH 0012260;%%
T.~Hahn, C.~Schappacher,
Comput.\ Phys.\ Commun.\  {\bf 143}, 54 (2002);
%%CITATION = HEP-PH 0105349;%%
J.~K\"ublbeck, M.~B\"ohm, A.~Denner,
Comput.\ Phys.\ Commun.\  {\bf 60}, 165 (1990).
%%CITATION = CPHCB,60,165;%%

\bibitem{formcalc}
 T.~Hahn and J.~I.~Illana,
  %``Extensions in FormCalc 5.3,''
  arXiv:0708.3652 [hep-ph].
  %%CITATION = ARXIV:0708.3652;%%
 T.~Hahn and J.~I.~Illana,
  %``Excursions into FeynArts and FormCalc,''
  Nucl.\ Phys.\ Proc.\ Suppl.\  {\bf 160} (2006) 101;
  %%CITATION = NUPHZ,160,101;%%
T.~Hahn, M.~Perez-Victoria,
Comput.\ Phys.\ Commun.\  {\bf 118}, 153 (1999);
%%CITATION = HEP-PH 9807565;%%
T.~Hahn and J.~I.~Illana,
  %``Extensions in FormCalc 5.3,''
  arXiv:0708.3652 [hep-ph].
  %%CITATION = ARXIV:0708.3652;%%
T.~Hahn,
  %``Automatic loop calculations with FeynArts, FormCalc, and LoopTools,''
  Nucl.\ Phys.\ Proc.\ Suppl.\  {\bf 89}, 231 (2000).
  %%CITATION = NUPHZ,89,231;%%

\bibitem{looptools}
G.~J.~van Oldenborgh,
Comput.\ Phys.\ Commun.\  {\bf 66}, 1 (1991);
%%CITATION = CPHCB,66,1;%%
T.~Hahn, Acta Phys.\ Polon.\ B {\bf 30}, 3469 (1999).
%%CITATION = HEP-PH 9910227;%%

\bibitem{cuba}
T.~Hahn,
  %``CUBA: A library for multidimensional numerical integration,''
  Comput.\ Phys.\ Commun.\  {\bf 168} (2005) 78
  [arXiv:hep-ph/0404043].
  %%CITATION = CPHCB,168,78;%%
T.~Hahn,
  %``The CUBA library,''
  Nucl.\ Instrum.\ Meth.\  A {\bf 559} (2006) 273
  [arXiv:hep-ph/0509016].
  %%CITATION = NUIMA,A559,273;%%

\bibitem{brucher}
A.~Stange, W.~J.~Marciano and S.~Willenbrock,
  %``Higgs Bosons At The Fermilab Tevatron,''
  Phys.\ Rev.\  D {\bf 49}, 1354 (1994)
  [arXiv:hep-ph/9309294];
  %%CITATION = PHRVA,D49,1354;%%
M.~A.~Diaz and T.~J.~Weiler,
  %``Decays of a fermiophobic Higgs,''
  arXiv:hep-ph/9401259.
  %%CITATION = HEP-PH/9401259;%%
 A.~Barroso, L.~Brucher and R.~Santos,
  %``Is there a light fermiophobic Higgs?,''
  Phys.\ Rev.\  D {\bf 60}, 035005 (1999)
  [arXiv:hep-ph/9901293].
  %%CITATION = PHRVA,D60,035005;%%
L.~Brucher and R.~Santos,
  %``Experimental signatures of fermiophobic Higgs bosons,''
  Eur.\ Phys.\ J.\  C {\bf 12}, 87 (2000)
  [arXiv:hep-ph/9907434].
  %%CITATION = EPHJA,C12,87;%%
 A.~G.~Akeroyd,
  %``Three-body decays of Higgs bosons at LEP2 and application to a hidden
  %fermiophobic Higgs,''
  Nucl.\ Phys.\  B {\bf 544}, 557 (1999)
  [arXiv:hep-ph/9806337].
  %%CITATION = NUPHA,B544,557;%%

\bibitem{LEPf}
 P.~Abreu {\it et al.}  [DELPHI Collaboration],
  %``Search for a fermiophobic Higgs at LEP 2,''
  Phys.\ Lett.\  B {\bf 507}, 89 (2001)
  [arXiv:hep-ex/0104025].
  %%CITATION = PHLTA,B507,89;%%

\bibitem{andrewdiaz}
A.~G.~Akeroyd, M.~A.~Diaz and F.~J.~Pacheco,
  %``Double fermiophobic Higgs boson production at the LHC and LC,''
  Phys.\ Rev.\  D {\bf 70}, 075002 (2004)
  [arXiv:hep-ph/0312231].
  %%CITATION = PHRVA,D70,075002;%%


\bibitem{maria} 
I.~F.~Ginzburg, M.~Krawczyk, P.~Osland,
%``Potential of photon collider in resolving SM-like scenarios,''
Nucl.\ Instrum.\ Meth.\ A {\bf 472}, 149 (2001).
%%CITATION = HEP-PH 0101229;%%

% Indirect information on triple coupling through radi.cor
\bibitem{indirect1} 
A.~Arhrib, W.~Hollik, S.~Penaranda and M.~Capdequi Peyranere,
  %``Higgs decays in the two Higgs doublet model: Large quantum effects in the
  %decoupling regime,''
  Phys.\ Lett.\  B {\bf 579} (2004) 361.
  %%CITATION = PHLTA,B579,361;%%


\bibitem{okada}
  S.~Kanemura, S.~Kiyoura, Y.~Okada, E.~Senaha and C.~P.~Yuan,
  %``New physics effect on the Higgs self-coupling,''
  Phys.\ Lett.\  B {\bf 558} (2003) 157
  [arXiv:hep-ph/0211308],
  %%CITATION = PHLTA,B558,157;%%
 S.~Kanemura, Y.~Okada, E.~Senaha and C.~P.~Yuan,
  %``Higgs coupling constants as a probe of new physics,''
  Phys.\ Rev.\  D {\bf 70} (2004) 115002
  [arXiv:hep-ph/0408364].
  %%CITATION = PHRVA,D70,115002;%%








\end{thebibliography}
\end{document}